\documentclass[11pt,a4paper]{article}
\pdfoutput=1
\usepackage{jheppub}

\newcommand{\be}{\begin{equation}}
\newcommand{\ee}{\end{equation}}
\newcommand{\bea}{\begin{eqnarray}}
\newcommand{\eea}{\end{eqnarray}}

\newcommand{\eqn}[1]{(\ref{#1})}

\newcommand{\mt}[1]{\textrm{\tiny #1}}
\def\nc {N_\mt{c}}

\newcommand{\sac}{\, , \qquad}

\newcommand{\uh}{u_\mt{H}}

\newcommand{\cf}{{\cal F}}
\newcommand{\cb}{{\cal B}}
\newcommand{\ch}{{\cal H}}
\newcommand{\cn}{{\cal N}}

\newcommand{\vp}{\varphi}
\newcommand{\qh}{\hat{q}}
\newcommand{\qiso}{\hat{q}_\mt{iso}}
\newcommand{\ql}{\hat q_\mt{L}}
\newcommand{\qp}{\hat q_\perp}

\title{Jet quenching in a strongly coupled anisotropic plasma}
\author[a,b]{Mariano Chernicoff,}
\author[a]{Daniel Fern\'andez,}
\author[a,c]{David Mateos,}
\author[d,e]{and Diego Trancanelli}
\affiliation[a]{Departament de F\'\i sica Fonamental \&  Institut de 
Ci\`encies del Cosmos (ICC), Universitat de Barcelona (UB), Mart\'{\i}  i
Franqu\`es 1, E-08028 Barcelona, Spain}  
\affiliation[b]{Department of Applied Mathematics and Theoretical Physics, Centre for Mathematical Sciences,
Wilberforce Road, Cambridge, CB3 0WA, UK}
\affiliation[c]{Instituci\'o Catalana de Recerca i Estudis Avan\c cats (ICREA),
Passeig Llu\'\i s Companys 23, E-08010, Barcelona, Spain} 
\affiliation[d]{Department of Physics, University of Wisconsin, Madison, WI
53706, USA}  
\affiliation[e]{Instituto de F\'\i sica, Universidade de S{\~a}o Paulo, 05314-970 S{\~a}o Paulo, Brazil}

\date{\today}

\abstract{The jet quenching parameter of an anisotropic plasma depends on the relative orientation between the anisotropic direction, the direction of motion of the parton, and the direction along which the momentum broadening is measured. We calculate the jet quenching parameter of an anisotropic, strongly coupled ${\cal N}=4$ plasma by means of its gravity dual. We present the results for arbitrary orientations and arbitrary values of the anisotropy. The anisotropic value can be larger or smaller than the isotropic one, and this depends on whether the comparison is made at equal temperatures or at equal entropy densities. We compare our results to analogous calculations for the real-world quark-gluon plasma and find agreement in some cases and disagreement in others.
}  
  

\emailAdd{M.Chernicoff@damtp.cam.ac.uk}
\emailAdd{daniel@ffn.ub.edu} 
\emailAdd{dmateos@icrea.cat} 
\emailAdd{dtrancan@fma.if.usp.br} 

\begin{document}

\begin{flushright}
DAMTP-2012-15 \\
ICCUB-12-098 \\
MAD-TH-12-02
\end{flushright}

\maketitle
\setlength{\parskip}{8pt}


\section{Introduction}
\label{intro}
A remarkable conclusion from the experiments at the Relativistic Heavy Ion Collider (RHIC)  \cite{rhic,rhic2} and at the Large Hadron Collider (LHC) \cite{lhc} is that the quark-gluon plasma (QGP) does not behave as a weakly coupled gas of quarks and gluons, but rather as a strongly coupled fluid \cite{fluid,fluid2}. This renders perturbative methods inapplicable in general. The lattice formulation of Quantum Chromodynamics (QCD) is also of limited utility, since for example it is not well suited for studying real-time phenomena. This has provided a strong motivation for understanding the dynamics of strongly coupled non-Abelian plasmas through the gauge/string duality \cite{duality,duality2,duality3} (see \cite{review} for a recent review of applications to the QGP). 

For a period of time $\tau_\mt{out}$ immediately after the collision, the system thus created is  anisotropic and far from equilibrium. After a time $\tau_\mt{iso} > \tau_\mt{out}$ the system becomes locally isotropic. It has been proposed than an intrinsically anisotropic hydrodynamical description can be used to describe the system at intermediate times $\tau_\mt{out} < \tau < \tau_\mt{iso}$ \cite{ani,ani2,ani3,ani4,ani5,ani6,ani7,ani8,ani9}. In this phase the plasma is assumed to have significantly unequal pressures in the longitudinal and transverse directions. The standard hydrodynamic description is a derivative expansion around equal pressures, and therefore it is not applicable in this regime. In contrast, the intrinsically anisotropic hydrodynamical description is a derivative expansion around an anisotropic state, and hence in this case the requirement that derivative corrections be small does not imply small pressure differences. In a real collision the degree of anisotropy will decrease with time, but for some purposes it is a good approximation to take it to be constant over an appropriate time scale. 

Motivated by these considerations, in this paper we will investigate the effect of an intrinsic anisotropy on the momentum broadening experienced by a fast parton moving through the plasma. For this purpose we will compute the jet quenching parameter for an ultra-relativistic quark propagating through an  anisotropic ${\cal N}=4$ super Yang-Mills plasma by means of its gravity dual \cite{prl,jhep}. As we will review below, the plasma is held in anisotropic equilibrium by an external force. The gravity solution possesses an anisotropic horizon, it is completely regular on and outside the horizon, and it is solidly embedded in type IIB string theory. For these reasons it provides an ideal toy model in which questions about  anisotropic effects at strong coupling can be addressed from first principles.

Previous calculations of the jet quenching parameter in the presence of anisotropy in the context of the gauge/gravity correspondence include \cite{sadeghi1,sadeghi2}. While this paper was being typewritten we received \cite{Giataganas:2012zy}, in which the jet quenching parameter along particular directions in the background of \cite{prl,jhep}  is studied in the limit of small anisotropy.

\section{Gravity solution}
\label{grav}
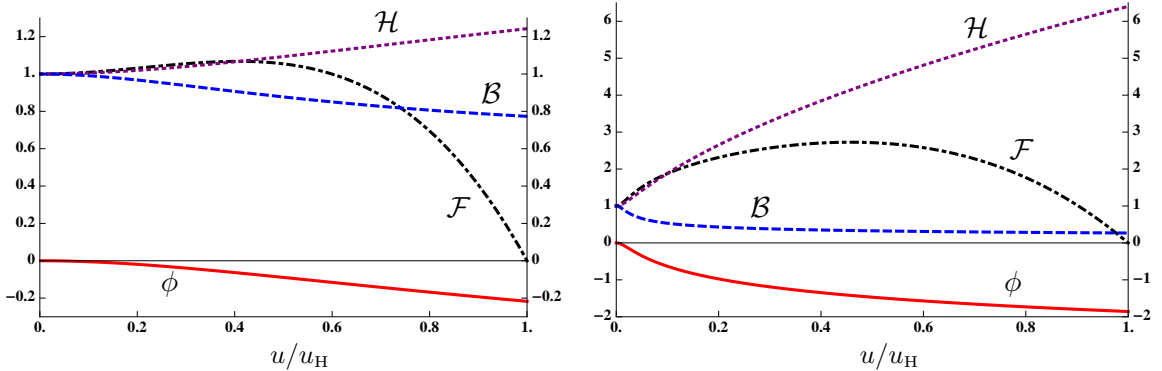
\begin{figure}[tb]
\begin{center}
\begin{tabular}{cc}
\includegraphics[scale=0.8]{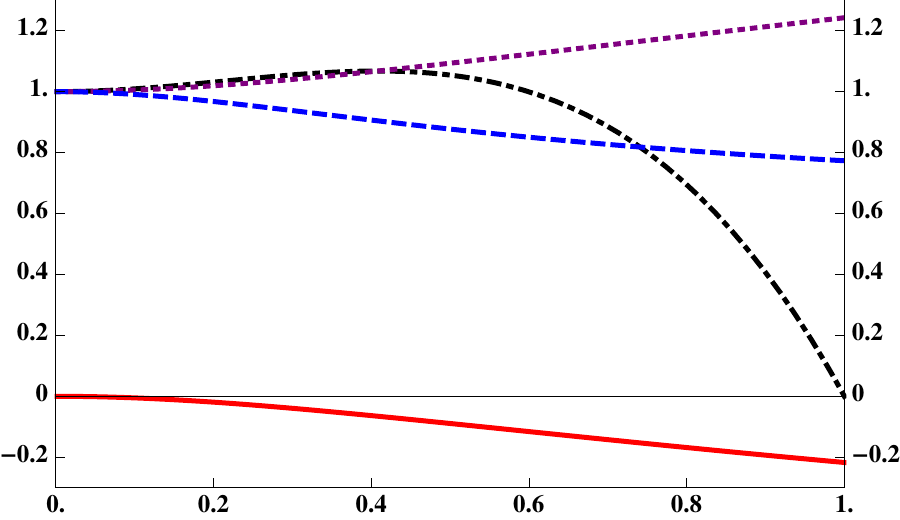}
\put(-109,-10){\small $u/\uh$}
\put(-109,-14){$$}
\put(-150,17){$\phi$}
\put(-70,115){$\ch$}
\put(-30,88){$\cb$}
\put(-43,45){$\cf$}
&
\includegraphics[scale=0.8]{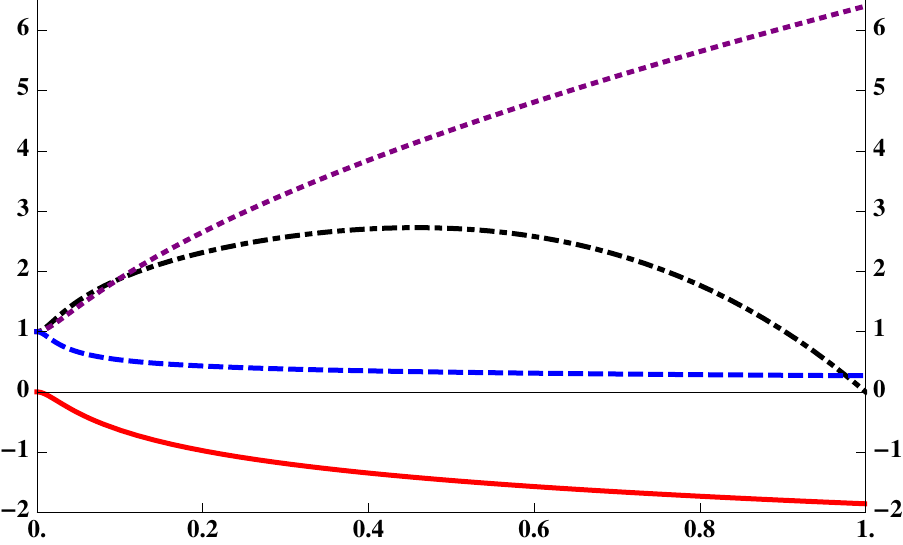}
\put(-109,-10){\small $u/\uh$}
\put(-109,-14){$$}
\put(-55,17){$\phi$}
\put(-70,113){$\ch$}
\put(-150,45){$\cb$}
\put(-53,67){$\cf$}
\end{tabular}
\caption{\small Metric functions for $a/T\simeq 4.4$ (left) and $a/T\simeq 86$ (right).
\label{plots}
}
 \end{center} 
 \end{figure}

The type IIB supergravity solution of \cite{prl,jhep} in the string frame takes the form
\bea
&&\hskip -.35cm 
ds^2 =  \frac{L^2}{u^2}
\left( -\cf \cb\, dt^2+dx^2+dy^2+ \ch dz^2 +\frac{ du^2}{\cf}\right) +
L^2 e^{\frac{1}{2}\phi} d\Omega_5^2, 
\,\,\,\,\,\,\label{sol1} \\
&& \hskip -.35cm \chi = az \sac \phi=\phi(u) \,,
\label{sol2}
\eea
where $\chi$ and $\phi$ are the axion and the dilaton, respectively, and $(t,x,y,z)$ are the gauge theory coordinates. 
Since there is rotational invariance in the $xy$-directions, we will refer to these as the transverse directions, and to $z$ as the longitudinal direction. $\cf, \cb$ and $\ch$ are functions of the holographic radial coordinate $u$ that were determined numerically in \cite{prl,jhep}. Their form for two values of $a/T$ is plotted in Fig.~\ref{plots}. 
The horizon lies at $u=\uh$, where $\cf=0$, and the boundary lies at $u=0$, where $\cf=\cb=\ch=1$ and $\phi=0$. The metric near the boundary asymptotes to $AdS_5 \times S^5$. Note that the axion is linear in the $z$-coordinate. The proportionality constant $a$ has dimensions of mass and is a measure of the anisotropy. The axion profile is dual in the gauge theory to a position-dependent theta parameter that depends linearly on $z$. This acts as an isotropy-breaking external source that forces the system into an anisotropic equilibrium state. 

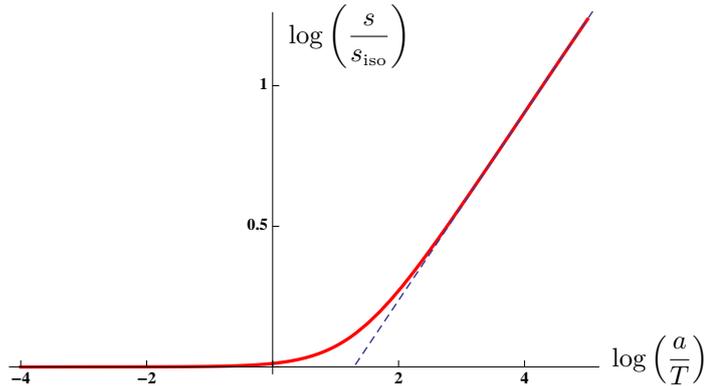
\begin{figure}[t!]
\begin{center}
\includegraphics[scale=0.85]{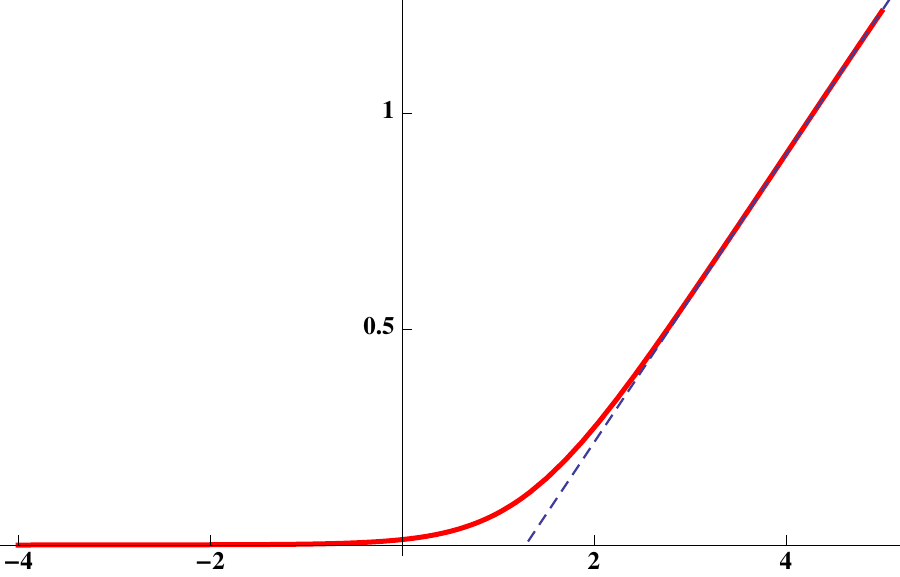}
 \begin{picture}(0,0)
   \put(1,8){\small{$\log \left( 
   {\displaystyle \frac{a}{T}} \right)$}}
    \put(-120,130){\small{$\log \left( 
    {\displaystyle \frac{s}{s_\mt{iso}}} \right)$}}
 \end{picture}
\caption{\small Log-log plot of the entropy density as a function of 
$a/T$, with $s_\mt{iso}$ defined as in eqn.~\eqn{siso}. The dashed blue line is a straight line with slope $1/3$.
\label{scalings}}
\end{center}
\end{figure}

If $a=0$ then the solution reduces to the isotropic black D3-brane solution dual to the isotropic 
$\cn=4$ theory at finite temperature. In this case
\be
\cb=\ch=1 \sac \chi=\phi=0 \sac \cf = 1-\frac{u^4}{\uh^4}
\sac \uh = \frac{1}{\pi T} 
\label{iso}
\ee
and the entropy density takes the form 
\be
s_\mt{iso}= \frac{\pi^2}{2} \nc^2 T^3 \,.
\label{siso}
\ee
Fig.~\ref{scalings} shows the entropy density of the anisotropic
plasma as a function of the dimensionless ratio $a/T$, normalized to the entropy density of the isotropic plasma at the same temperature. At small $a/T$ the entropy density scales as in the isotropic case, whereas at large $a/T$ it scales as \cite{ALT,prl,jhep}
\be
s = c_\mt{ent} \nc^2 a^{1/3} T^{8/3} \,,
\label{larges}
\ee
where $c_\mt{ent}\simeq 3.21$. The transition between the two behaviours takes place approximately around $a/T \simeq 3.7$.  

A feature of the solution \eqn{sol2} that played an important role in the analysis of \cite{prl,jhep} is the presence of a conformal anomaly. Its origin lies in the fact that diffeomorphism invariance in the radial direction $u$ gets broken in the process of renormalization of the on-shell supergravity action. In the gauge theory this means that scale invariance is broken by the renormalization process. One manifestation of the anomaly is the fact that, unlike the entropy density, other thermodynamic quantities do not depend solely on the ratio $a/T$ but on $a$ and $T$ separately. This will not be the case for the jet quenching parameter, which as we will see takes the form $\hat{q}(a,T) = T^3 f(a/T)$.  


\section{Jet quenching parameter}
In this section we will calculate the jet quenching parameter  $\qh$ for an ultra-relativistic quark following the prescription of Refs.~\cite{liu2,liu1,redux}. This instructs us to consider the worldsheet of a string whose endpoints move at the speed of light  along a given boundary direction and are separated a small distance $\ell$ along an orthogonal direction. The former is the direction of motion of the quark, and the latter is the direction along which the momentum broadening takes place. In the presence of  anisotropy the jet quenching parameter depends on how these directions are oriented with respect to the longitudinal and transverse directions in the plasma.  Recall that there is rotational symmetry in the $xy$-directions but not in the $z$-direction. In the context of a heavy ion collision $z$ would correspond to the longitudinal, beam direction, and $x,y$ to the directions in the transverse plane. 
\begin{figure}[t!]
\begin{center}
\includegraphics[scale=.9]{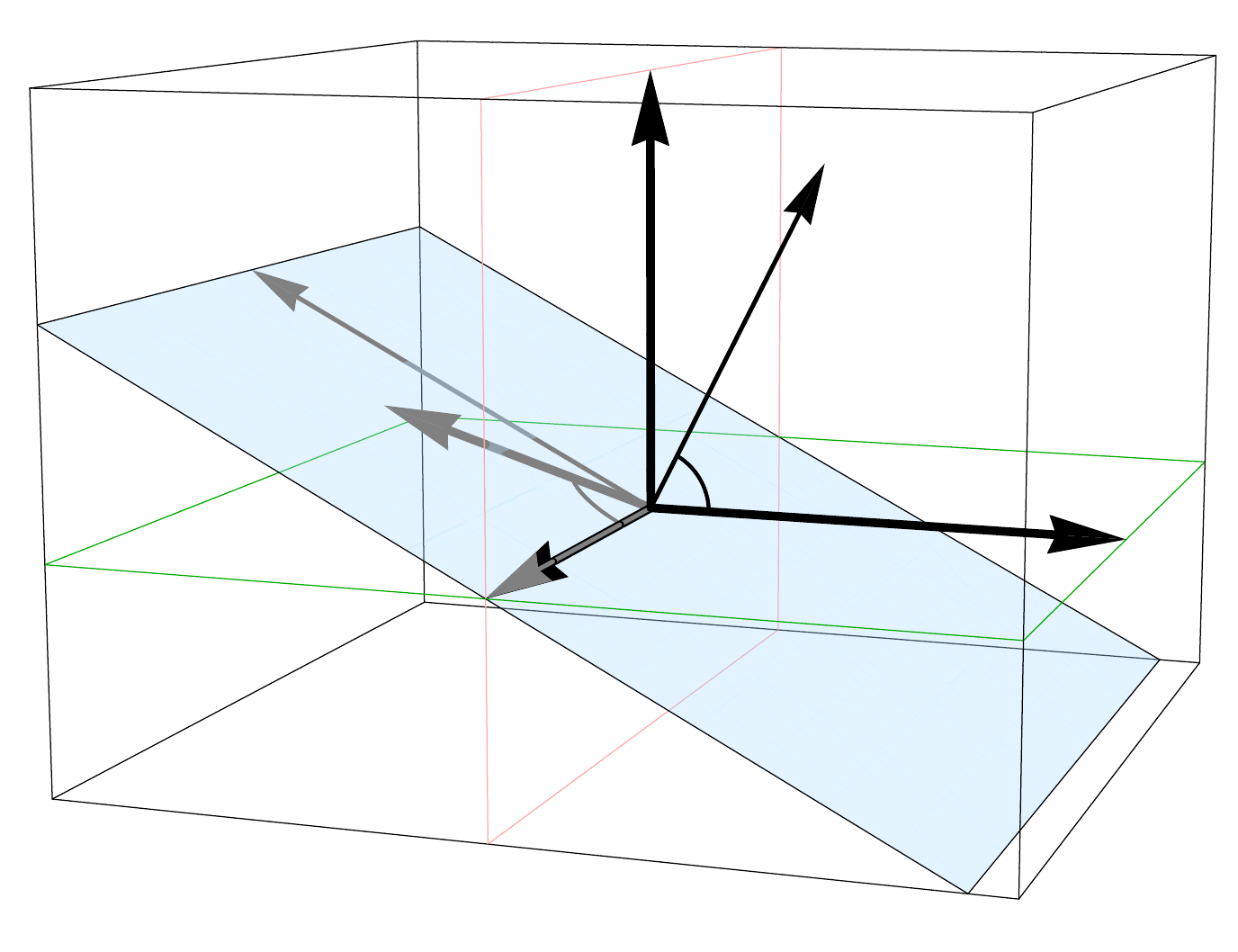}
 \begin{picture}(0,0)
   \put(-190,233){\Large $x$}
   \put(-120,217){{\large $Z$}}
       \put(-50,127){{\Large $z$}}
    \put(-297,182){\large $X$}
    \put(-255,88){{\Large $y$}\, = \,{\large $Y$}}
    \put(-155,137){$\theta$}
    \put(-202,125){$\vp$}
         \put(-268,153){$\Delta p$}
 \end{picture}
\caption{Relative orientation between the anisotropic direction $z$, the direction of motion of the quark $Z$, and the direction in which the momentum broadening is measured $\Delta p$. In the context of a heavy ion collision $z$ would correspond to the longitudinal direction along the beam, and $x,y$ to the directions in the transverse plane. The direction of motion lies in the $xz$-plane at an angle $\theta$ with respect to the $z$-axis. The momentum broadening takes place in any direction in the $XY$-plane orthogonal to $Z$. $X$ lies within the $xz$-plane, whereas $y=Y$ is orthogonal to it. The angle in the 
$XY$-plane between the $Y$ axis and the direction of the momentum broadening is $\vp$.
}
\label{geom}
\end{center}
\end{figure}
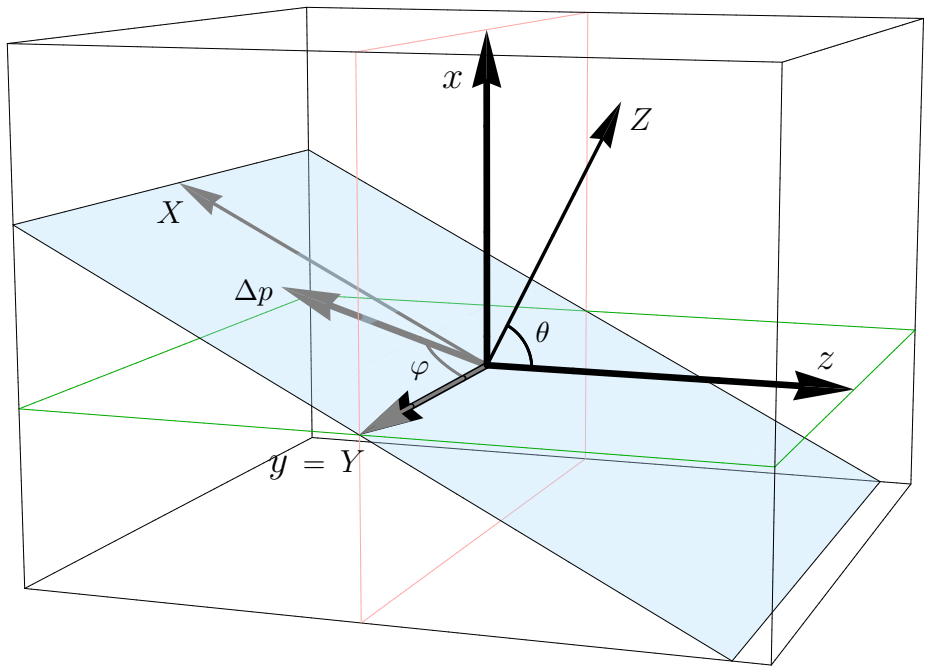
Given the rotational symmetry in this plane, we will assume without loss of generality that the the direction of motion is contained in the $xz$-plane, and  we will denote it by $Z$ (see Fig.~\ref{geom}). We call $\theta$ the angle between this direction and the $z$-axis. The two independent orthogonal directions to $Z$ can then be chosen so that one, which we denote by $X$, lies within the $xz$-plane, and the other one, $Y$, coincides with the $y$-axis. We denote as $\vp$ the polar angle in the $XY$-plane with respect to the $Y$-axis. The $XYZ$-coordinate system is obtained from the $xyz$-coordinate system by a rotation of angle $\theta$ around the $y=Y$ axis, as described by eqn.~\eqn{change1} below. We will determine the jet quenching parameter associated to momentum broadening in an arbitrary direction in the $XY$-plane, and we will refer to it as $\qh_{\theta,\vp}$ in order to emphasize that it depends on the two angles defined above. 

Now recall that $\qh$ is the average momentum squared acquired by the quark after traveling through the medium a unit distance \cite{Baier}. If we call $p_\vp$ the component of the momentum in the direction within the $XY$-plane specified by $\vp$, then clearly 
\be
p_\vp = p_Y \cos \vp + p_X \sin \vp \,.
\ee
Squaring and taking an average we obtain 
\be
\langle \Delta p_\vp^2 \rangle = \cos^2 \vp \langle \Delta p_Y^2 \rangle + \sin^2 \vp \langle \Delta p_X^2 \rangle \,,
\ee
where we used the fact that $\langle \Delta p_Y \Delta p_X \rangle = 0$ given the symmetry under $Y\to -Y$. Rewritten in terms of the corresponding jet quenching parameters this becomes 
\be
\qh_{\theta,\vp} =  \qh_{\theta,0} \, \cos^2 \vp  + 
\qh_{\theta,\pi/2} \, \sin^2 \vp \,.
\label{relrel}
\ee
We will see that the gravity calculation reproduces this relation. 

Rather than starting with the most general case, for pedagogical reasons we will first study two particular cases corresponding to motion along the longitudinal direction and motion contained within the transverse plane. The general case will be discussed in Sec.~\ref{3}.


\subsection{Motion along the longitudinal direction}
\label{1}
This case corresponds to $\theta=0$ and is the simplest one because the momentum broadening takes place in the transverse $xy$-plane, which is rotationally symmetric. In particular, this means that the result is independent of $\vp$, since $ \qh_{0,0}= \qh_{0,\pi/2}$. In the context of heavy ion collisions, this case corresponds to motion of the parton along the beam direction.

It is convenient to carry out the calculation using
the light cone coordinates
\begin{equation}
z^{\pm}=\frac{t\pm z}{\sqrt{2}}\,.
\end{equation}
Ignoring the sphere part, which will play no role in the following, the metric (\ref{sol1}) reads
\begin{equation}
ds^2=\frac{L^2}{u^2}\left[ \frac{1}{2}(\ch-\cf\cb)(dz^{+})^2+\frac{1}{2}
(\ch-\cf\cb)(dz^{-})^2-(\ch+\cf\cb)dz^{+}dz^{-}+dx^2+dy^2+\frac{du^2}{\cf}\right]\,.
\end{equation}
We consider a quark moving along $z^-$. Given the symmetry in the $xy$-plane we set $y=0$ without loss of generality. We then fix the static gauge by identifying 
$(z^-,x)=(\tau,\sigma)$ and  specify the string embedding through one function $u=u(x)$ subject to the boundary condition that $u(\pm \ell/2)=0$. 
Under these circumstances the Nambu-Goto action  
\begin{equation}
S=-\frac{1}{2\pi\alpha'} \int d\tau d\sigma 
\sqrt{-\det g_\mt{ind}}
\label{ng}
\end{equation}
takes the  form
\begin{equation}
\label{jetaction}  
S=2 i \frac{L^2}{2\pi\alpha'}\int{dz^{-}\int_0^{\ell/2}dx \,
\frac{1}{u^2} \sqrt{\frac{1}{2}
\left( \ch-\cf\cb \right) \left( 1+\frac{u'^2}{\cf} \right)}} \,,
\end{equation}
where the factor of 2 comes from  the fact that  the integral over $x$  covers only one half of the string, and $u'=du/dx$. Note that the action is imaginary because the string worldsheet is spacelike, as expected in order for the jet quenching parameter to be real \cite{liu1} (see also \cite{review} for an extensive discussion). The fact that the Lagrangian does not depend on $x$ explicitly leads to a conserved quantity $\Pi_x$ and to the first-order equation
\begin{equation}
\label{jeteomu}
u'^2=\frac{\cf}{2 \Pi_x^2 u^4} 
\left[ (\ch-\cf\cb) - 2 \Pi_x^2 u^4\right]\,.
\end{equation}
The turn-around point for the string is defined by $u'=0$. The prescription for computing the jet quenching parameter instructs us to work in the limit $\ell \to 0$. As we will see below, this corresponds to the limit  $\Pi_x \to 0$. In this case it is clear that the term inside the square brackets is positive. This follows from the fact that $\ch$ ($\cf \cb$) is monotonically increasing (decreasing) from the boundary to the horizon, and that near the boundary $\ch - \cf \cb$ scales as $a^2 u^2/4$. 
 
We thus see that in the limit of interest the string descends all the way into the bulk and turns around precisely at the black hole horizon, as in the isotropic case \cite{liu2,liu1}. As explained in \cite{redux}, the string worldsheet must have this property in order to be dual to a gauge theory Wilson loop with the  operator ordering required for the extraction of the jet quenching parameter. The reason is that this ordering can be implemented by thinking of the time coordinate $t$ as a complex coordinate and requiring the worldlines of the quark and the antiquark to  lie on the $\mbox{Im}\, t=0$ and $\mbox{Im}\, t=-i\epsilon$ slices, respectively. In the black hole geometry \eqn{sol2} 
$\mbox{Im}\, t$ is periodic with period $1/T$ and these two slices only meet at the horizon, irrespectively of whether $a=0$ or $a\neq 0$. Therefore the string must descend from the boundary to the horizon on the (say) $\mbox{Im}\, t=0$ slice, turn around, and return to the boundary on the $\mbox{Im}\, t=-i\epsilon$ slice. However, since the metric on these two slices is identical, the resulting string action is the same as that of a horizon-touching string worldsheet that lies entirely on a single slice, and which is dual to a Wilson loop with a different operator ordering. This is the reason why the subtlety identified in Ref.~\cite{redux} did not change the isotropic result of Refs.~\cite{liu2,liu1}, which considered a single slice. Exactly the same equivalence applies in our anisotropic case, since all the string worldsheets that we will consider turn around precisely at the horizon. For this reason in what follows we will simply use the prescription from \cite{liu2,liu1}.

Integrating equation (\ref{jeteomu}) we obtain half the separation between the two endpoints of the string along the spatial side of the Wilson loop:
\begin{equation}\label{jetL}
\frac{\ell}{2} = \sqrt{2} \, \Pi_x\int^{\uh}_0{du\frac{u^2}
{\sqrt{\cf} \sqrt{ (\ch-\cf\cb) - 2 \Pi^2_x u^4}}}\,.
\end{equation} 
Note that, as anticipated above, $\ell \to 0$ as $\Pi_x \to 0$, and in this limit we have 
\be
\ell = 2 \sqrt{2} \, \Pi_x {\cal I}_x + {\cal O}(\Pi_x^2)
\label{rel}
\ee
with
\be
{\cal I}_x \equiv \int^{\uh}_0{du\frac{u^2}
{\sqrt{\cf} \sqrt{ \ch-\cf\cb}}}
\ee
a convergent integral.

To compute the jet quenching parameter we need to evaluate 
the on-shell action (\ref{jetaction}) on the solution  (\ref{jeteomu}). After changing the integration variable from $x$ to $u$ the result is
\begin{equation}
S=i \frac{\sqrt{\lambda} \, L^{-}}{\sqrt{2}\, \pi}\int^{\uh}_{0}du\frac{(\ch-\cf\cb)}
{u^2 \sqrt{\cf} \sqrt{(\ch-\cf\cb)- 2\Pi^2_xu^4}} \,,
\end{equation}
where $L^{-}$ is the long side of the Wilson loop. This action diverges due to the integration near $u=0$. This can be seen by expanding in powers of $\Pi_x$,
\begin{equation}
S=i \frac{\sqrt{\lambda} \, L^{-}}{\sqrt{2}\, \pi} 
\int^{\uh}_{0}du\frac{\sqrt{\ch-\cf\cb}}{u^2 \sqrt{\cf}}
+ i \frac{\sqrt{\lambda} \, L^{-} \ell^2}{8 \sqrt{2}\, \pi \, {\cal I}_x}
+ {\cal O}(\ell^4) \,,
\end{equation}
where we have used the relation \eqn{rel}. All terms of order $\ell^2$ and higher are finite, whereas the first, $\ell$-independent term diverges as $\log u$. This term can be renormalized away using several methods, including subtraction of the action of two disconnected strings \cite{liu2,liu1} or addition to the string action of a counterterm proportional to  $\log u \int d\tau \sqrt{\gamma}$, where $\gamma$ is the induced worldline metric on a constant-$\sigma$ slice of the string worldsheet. The logarithm  in this counterterm illustrates the fact that the renormalized string action is sensitive to the conformal anomaly in the gauge theory \cite{prl,jhep}. However, the jet quenching parameter is given by the finite $\ell^2$-term, whose extraction does not require any renormalization. It thus follows that $\hat q$ is insensitive to the presence of the anomaly, as anticipated in the Introduction. Using the prescription from \cite{liu2,liu1}, 
\begin{equation}
e^{i 2 S}=\langle W^A(C) \rangle=\exp\left[
-\frac{L^{-}\ell^2}{4\sqrt{2}}\,\hat{q} \right] \,,
\label{pres}
\end{equation}
where $S$ denotes the finite part of the action, we finally arrive at 
\begin{equation}
\label{jetqx}
\hat{q}_z\equiv \qh_{0,\vp} = \frac{\sqrt{\lambda}}{\pi \mathcal{I}_x} \,,
\end{equation}
where the subscript in $\hat q_z$ reminds us of the direction of motion of the quark. Eqn.~\eqn{jetqx} reduces to the correct result in the isotropic limit. In this case, using \eqn{iso}, we see that 
\be
{\cal I}_x =  \uh^2 \int^{\uh}_0{du\frac{1}
{\sqrt{1-u^4/\uh^4}}} = 
\frac{1}{\pi^3 T^3} \frac{\sqrt{\pi} \, 
\Gamma \left(\frac{5}{4} \right)}{\Gamma \left(\frac{3}{4} \right)}\,.
\ee
Substituting into \eqn{jetqx} we reproduce the isotropic result 
 \cite{liu2,liu1}
\be
\qiso (T) = \frac{\pi^{3/2} \, \Gamma \left(\frac{3}{4} \right)}
{\Gamma \left(\frac{5}{4} \right)} \sqrt{\lambda} \, T^3 \,.
\label{qisoT}
\ee
For later purposes it is useful to rewrite this in terms of the entropy density \eqn{siso} as 
\be
\qiso (s) = \frac{2 \Gamma \left(\frac{3}{4} \right)}
{\sqrt{\pi} \, \Gamma \left(\frac{5}{4} \right)} 
\sqrt{\lambda} \, \frac{s}{\nc^2} \,.
\label{qisos}
\ee

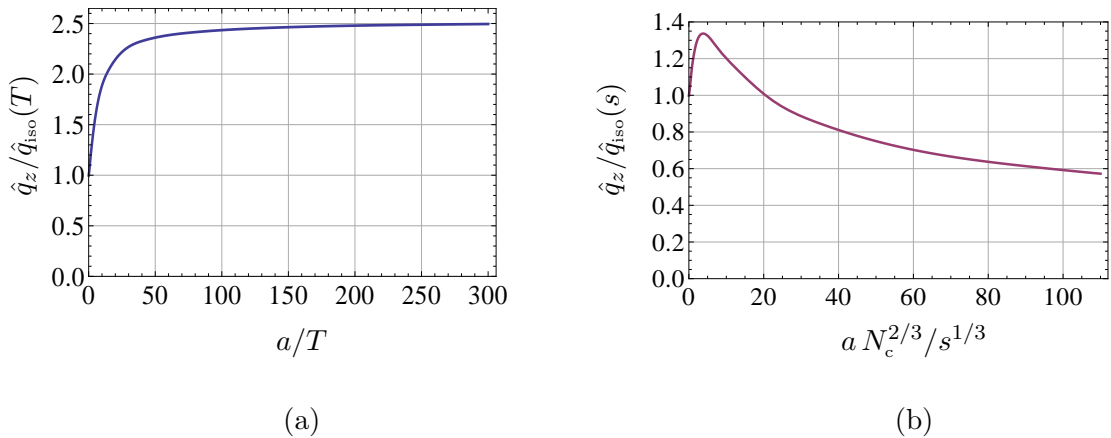
\begin{figure}[htb]
\begin{center}
\begin{tabular}{cc}
\setlength{\unitlength}{1cm}
\includegraphics[width=6cm,height=4cm]{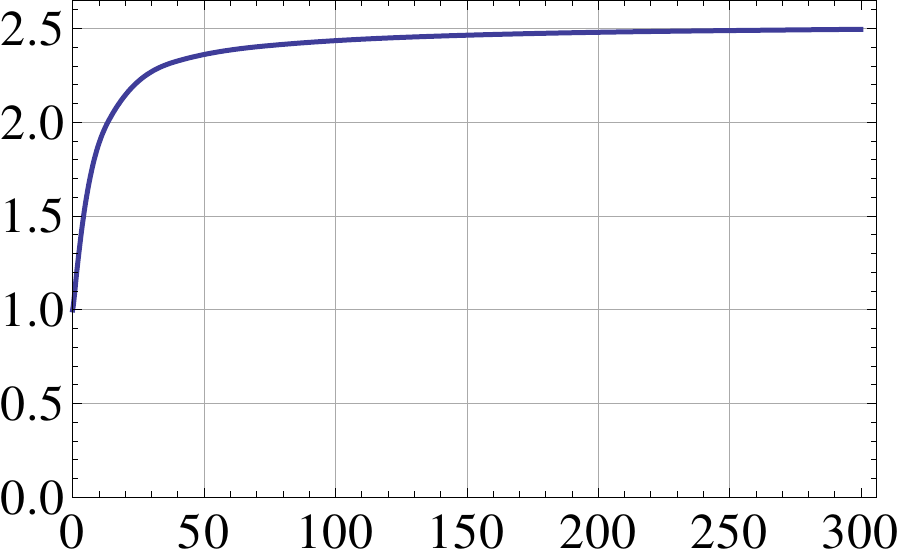}
\,\,\, \,\,\,\,  & \qquad
\includegraphics[width=6cm,height=4cm]{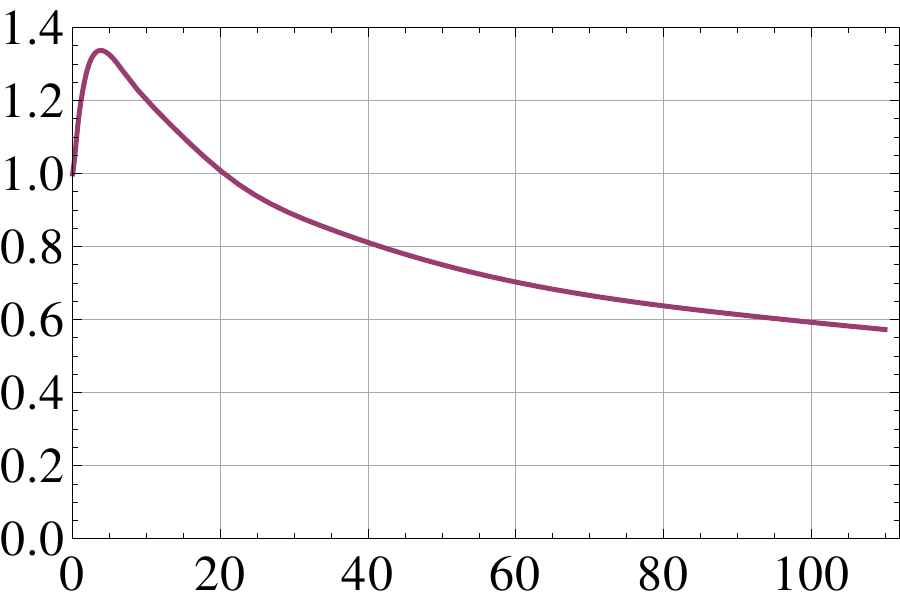}
 \begin{picture}(0,0)
  \put(-415,43){\rotatebox{90}{$\qh_z/\qh_\mt{iso}(T)$}} 
  \put(-195,43){\rotatebox{90}{$\qh_z/\qh_\mt{iso}(s)$}} 
        \end{picture}
    \\ 
  $a/T$ & \qquad\qquad $a \, \nc^{2/3} /s^{1/3}$ \\
  & \\
     (a) & \qquad\qquad (b)
     \end{tabular}
\caption{Jet quenching parameter for a quark moving along the longitudinal $z$-direction as a function of the anisotropy. 
$\qh_z=\qh_{0,\vp}$ and $a$ are plotted in the appropriate units to facilitate comparison with the isotropic result for a plasma at the same temperature (a), or at the same entropy density (b). The isotropic result is given in eqs.~\eqn{qisoT} and \eqn{qisos}.
}\label{qzT}
\end{center}
\end{figure}

Since for general $a$ the metric functions in \eqn{sol2} are only known numerically, we have numerically determined  $\qh_z$ as a function of the magnitude of the anisotropy $a$ measured in units of the temperature or in units of the  entropy density (see Fig.~\ref{qzT}). The reason for working with both  is that we wish to compare the jet quenching in the anisotropic plasma to that in the isotropic plasma, and this can be done at least in two different ways: the two plasmas can be taken to have the same temperatures but different entropy densities, or the same entropy densities but different temperatures. 


\subsection{Motion in the transverse plane}
\label{2}
Given the rotational symmetry in the $xy$-plane, we will choose the direction of motion to be the $x$-direction. Thus this case corresponds to $\theta=\pi/2$ in the parametrization of Fig.~\ref{geom}. Since there is no symmetry between the $y$ and $z$ directions, in this case the result will depend on $\vp$. 

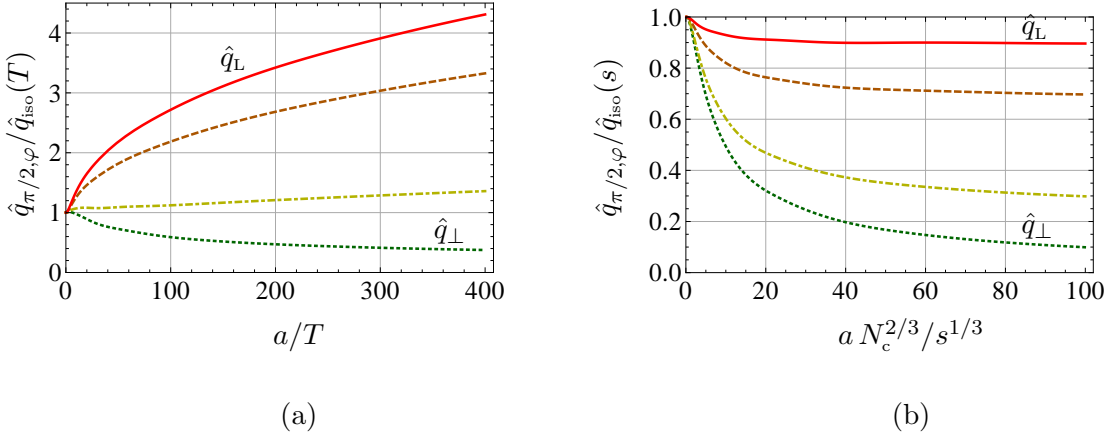
\begin{figure}[htb]
\begin{center}
\begin{tabular}{cc}
\setlength{\unitlength}{1cm}
\includegraphics[width=6cm,height=4cm]{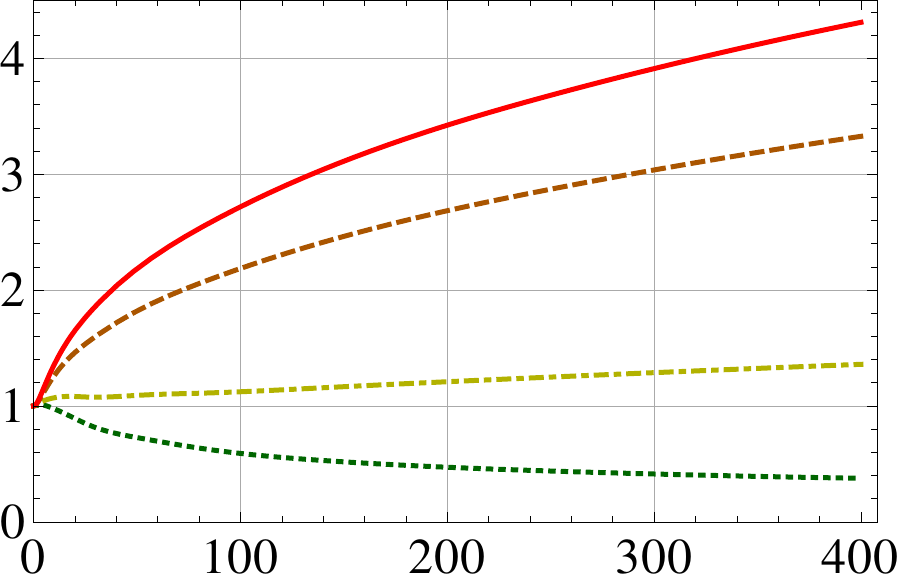}
\,\,\, \,\,\,\,  & \qquad
\includegraphics[width=6cm,height=4cm]{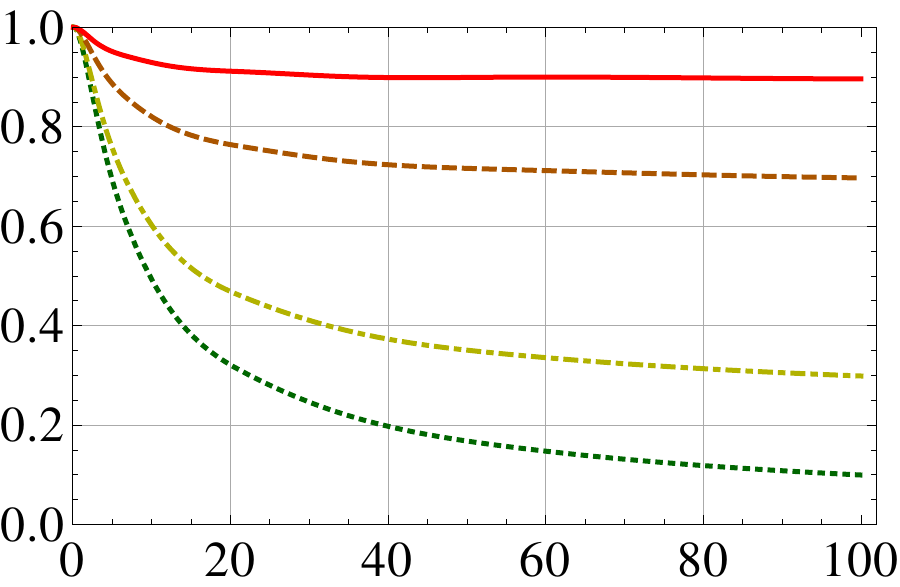}
 \begin{picture}(0,0)
  \put(-415,33){\rotatebox{90}
  {$\hat q_{\pi/2, \vp}/\qh_\mt{iso}(T)$}} 
  \put(-195,33){\rotatebox{90}
  {$\hat q_{\pi/2, \vp}/\qh_\mt{iso}(s)$}}
        \end{picture}
         \put(-335,90){$\ql$}
         \put(-255,25){$\qp$}
          \put(-35,102){$\ql$}
         \put(-35,27){$\qp$}
    \\ 
  $a/T$ & \qquad\qquad $a \, \nc^{2/3} /s^{1/3}$ \\
  & \\
     (a) & \qquad\qquad (b)
     \end{tabular} 
\caption{Jet quenching parameter $\hat q_{\pi/2, \vp}$ associated to momentum broadening in the $yz$-plane for a quark moving along the transverse direction $x$. The direction in the $yz$-plane lies at  an angle (from top to bottom) $\vp=\pi/2, \pi/3, \pi/6, 0$ 
with respect to the $y$-axis (see Fig.~\ref{geom}). The top (bottom) curves correspond to momentum broadening along the longitudinal (transverse) direction. $\hat q_{\pi/2, \vp}$ and $a$ are plotted in the appropriate units to facilitate comparison with the isotropic result for a plasma at the same temperature (a), or at the same entropy density (b). The isotropic result is given in eqs.~\eqn{qisoT} and \eqn{qisos}.
}\label{qytetaT}
\end{center}
\end{figure}

As in the previous example, it is convenient to work with adapted coordinates 
\begin{equation}
x^{\pm}=\frac{t\pm x}{\sqrt{2}} \,,
\end{equation}
in terms of which the metric takes the form
\begin{equation}
ds^2=\frac{L^2}{u^2}\left[
\frac{1}{2}(1-\cf\cb)(dx^{+})^2+\frac{1}{2}
(1-\cf\cb)(dx^{-})^2-(1+\cf\cb)dx^{+}dx^{-}+dy^2+\ch
dz^2+\frac{du^2}{\cf}\right]\,.
\end{equation}
In this case we choose the static gauge $(\tau,\sigma)=(x^{-},u)$, set $x^{+}=\mbox{const.}$, and specify the string projection in the $yz$-plane as 
\be
y \to \cos\vp \, y(u) \sac z \to \sin \vp \,  z(u) \,.
\ee 
Under these circumstances, the Nambu-Goto action \eqn{ng} becomes
\begin{equation}
S=2 i \frac{L^2}{2\pi\alpha'} \int{dx^{-} \int_0^{\uh}du
\frac{1}{u^2}\sqrt{\frac{1}{2}
\left( 1-\cf\cb \right)
\left( \frac{1}{\cf}+y'^2\cos^2\vp
+\ch z'^2\sin^2\vp \right)}} \,,
\label{jetactionxz}
\end{equation}
where the primes denote differentiation with respect to $u$ and the overall factor of 2 comes from the two branches of the string.
 We now follow the procedure in the previous section to obtain the jet quenching parameter. Since the Lagrangian does not depend on $y,z$ explicitly we find that
\be
\label{lx}
y'= \frac{\sqrt{2\ch} \,u^2\, \Pi_y }
{\sqrt{\cf} \sqrt{\ch \left( 1-\cf \cb \right) 
-2u^4 \left( \ch\, \Pi^2_y \cos^2\vp
+\Pi^2_z\sin^2\vp \right)}}
\ee
and
\be
\label{lz}
z'= \frac{\sqrt{2} \,u^2 \, \Pi_z}{\sqrt{\ch\cf}
{\sqrt{\ch \left( 1-\cf \cb \right) 
-2u^4 \left( \ch\, \Pi^2_y \cos^2\vp
+\Pi^2_z\sin^2\vp \right)}}} \,,
\ee
where $\Pi_y$ and $\Pi_z$ are conserved quantities (into which some factors of $\cos \vp$ and $\sin\vp$ have been absorbed).  An argument analogous to that in Sec.~\ref{1} shows that the denominators in these expressions only vanish at the horizon in the small-$\Pi$ limit. By integrating these equations we obtain the  separation between the two endpoints of the string. As in the previous section we will be interested in the limit $\Pi_y, \Pi_z \to 0$, so we work to  lowest order in these quantities: 
\be
\label{pepe}
\ell =2 \sqrt{2}\, \Pi_y \, \mathcal{I}_{xy} 
+ {\cal O} \left( \Pi^2 \right) \sac
\ell  = 2 \sqrt{2}\, \Pi_z \, \mathcal{I}_{xz} 
+ {\cal O} \left( \Pi^2 \right) \,,
\ee  
with 
\be
\label{ly}
\mathcal{I}_{xy} \equiv \int^{\uh}_0 du
\frac{u^2}{\sqrt{\cf(1-\cf \cb)}} 
\sac
\mathcal{I}_{xz}\equiv 
\int^{\uh}_0 du \frac{u^2}{\ch\sqrt{\cf(1-\cf \cb)}} 
\ee  
convergent integrals. Substituting the solution \eqn{lx}-\eqn{lz} into the action \eqn{jetactionxz}, expanding in powers of $\Pi$ and keeping only the term of order $\Pi^2$ we obtain
\be
\label{osactioyteta}
S=\frac{i \sqrt{\lambda}L^{-}}{\sqrt{2} \, \pi}
\int^{\uh}_0 du\left(
\frac{u^2\, \Pi^2_y\, \cos^2 \vp}{\sqrt{\cf(1-\cf \cb)}}+
\frac{u^2\, \Pi^2_z\, \sin^2 \vp}{\ch\sqrt{\cf(1-\cf \cb)}} \right) \,.
\ee
Using \eqn{pepe} and \eqn{ly} the action becomes
\be
S=\frac{i \sqrt{\lambda}L^{-} \ell^2}{8 \sqrt{2} \, \pi}
\left( \frac{\cos^2\vp}{
\mathcal{I}_{xy}}+\frac{\sin^2\vp}{\mathcal{I}_{xz}}
\right) \,,
\ee
so applying the prescription \eqn{pres} and defining 
\be
\hat{q}_\perp \equiv  \hat{q}_{\pi/2,0}=
\frac{\sqrt{\lambda}}{\pi \mathcal{I}_{yx}} \sac 
\hat{q}_\mt{L} \equiv  \hat{q}_{\pi/2,\pi/2}=
 \frac{\sqrt{\lambda}}{\pi \mathcal{I}_{yz}} 
\ee
we finally arrive at 
\be\label{jetqyteta}
\hat q_{\pi/2, \vp} =\hat{q}_\perp \cos^2\vp
+\hat{q}_\mt{L} \sin^2\vp \,.
\ee
This is a particular case of the relation \eqn{relrel} anticipated above. In Figure \ref{qytetaT} we have plotted this result  for  several values of $\vp$. 

 
\subsection{Arbitrary motion}
\label{3}
We now consider an arbitrary motion within the $xz$-plane, as explained in Fig.\ref{geom}. For this purpose we first define rotated coordinates $X,\, Z$ through
\bea
z&=& Z \cos \theta - X \sin \theta  \,, \nonumber \\
x &=& Z \sin \theta + X \cos \theta \,, \nonumber \\
y &=& Y \,,
\label{change1}
\eea
and then we go to light-cone coordinates by setting
\be
t = \frac{Z^- + Z^+}{\sqrt{2}} \,, \qquad 
Z = \frac{Z^- - Z^+}{\sqrt{2}} \,.
\label{change2}
\ee
Recall that $Z$ is the direction of motion. We thus fix the static gauge $Z^- =\tau, \, u=\sigma$, and seek a solution for the string embedding parametrized as 
\be
Z^+=Z^+(u) \sac X \to X(u) \sin \vp  \sac Y \to Y(u) \cos \vp \,.
\ee
With this choice $\vp$ is the polar angle in the plane orthogonal to $Z$ between the direction of momentum broadening and the $Y$-axis. Note that we must allow for a non-constant embedding in the $Z^+$-direction in order to find a  solution.
 
\begin{figure}[h!!!!!]
\begin{center}
\begin{tabular}{cc}
\setlength{\unitlength}{1cm}
\includegraphics[width=6cm,height=4cm]{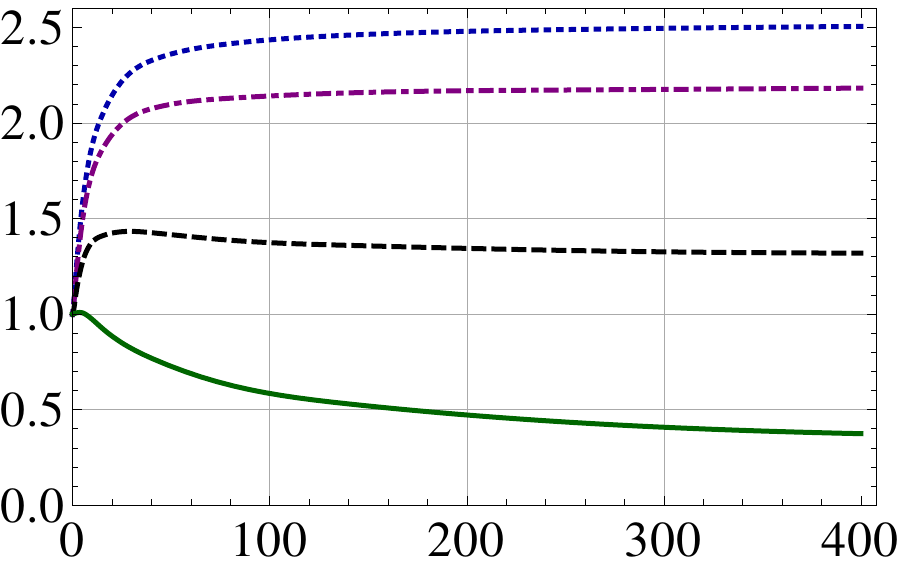}  \,\,\, \,\,\,\, 
& \qquad 
\includegraphics[width=6cm,height=4cm]{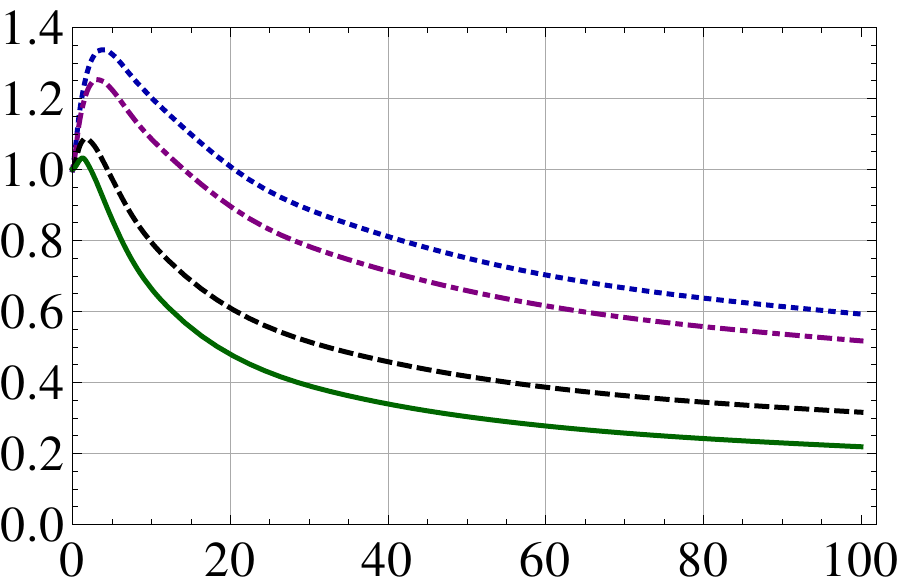}
 \begin{picture}(0,0)
  \put(-415,43){\rotatebox{90}
  {$\hat{q}_{\theta,0}/\qh_\mt{iso}(T)$}} 
  \put(-195,43){\rotatebox{90}
  {$\hat{q}_{\theta,0}/\qh_\mt{iso}(s)$}}
        \end{picture}
       \\ 
  $a/T$ & \qquad\qquad $a \, \nc^{2/3} /s^{1/3}$ \\
  & \\
     (a) & \qquad\qquad (b)  \\
    \, & \, \\
     \includegraphics[width=6cm,height=4cm]{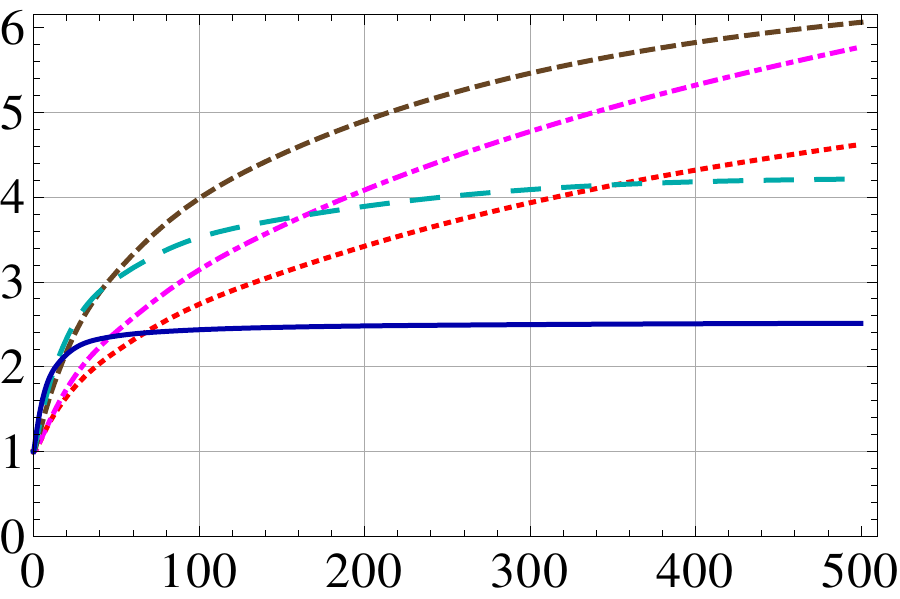}
 \,\,\, \,\,\,\, 
& \qquad 
\includegraphics[width=6cm,height=4cm]{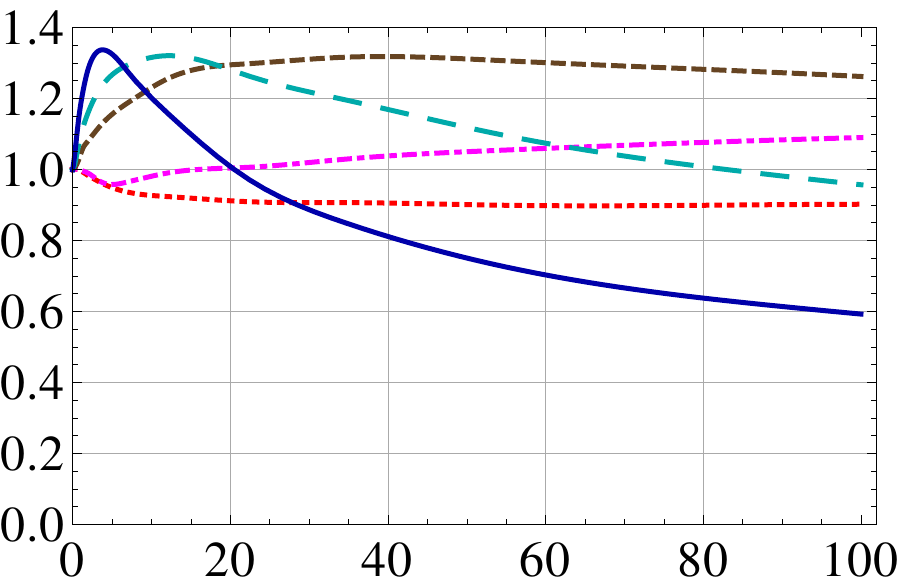}
 \begin{picture}(0,0)
  \put(-415,43){\rotatebox{90}
  {$\hat{q}_{\theta,\frac{\pi}{2}}/\qh_\mt{iso}(T)$}} 
  \put(-195,43){\rotatebox{90}
  {$\hat{q}_{\theta,\frac{\pi}{2}}/\qh_\mt{iso}(s)$}}
        \end{picture}
       \\ 
  $a/T$ & \qquad\qquad $a \, \nc^{2/3} /s^{1/3}$ \\
  & \\
     (c) & \qquad\qquad (d)
     \end{tabular}
\caption{
Jet quenching parameter for a quark moving along an arbitrary direction in the $xz$-plane, associated to momentum broadening along the transverse $y$-direction (top) or within the $xz$-plane (bottom). In (a) and (b) the angle between the direction of motion and the longitudinal $z$-direction is (from top to bottom) $\theta=0, \pi/6, \pi/3, \pi/2$, whereas the correspondence in (c) and (d) is $\theta=5\pi/12$ (brown, dashed), $49\pi/100$ (magenta, dotted-dashed), $\pi/2$ (red, dotted), $\pi/3$ (cyan, coarsely dashed), and $0$ (blue, continuous). $\qh_{\theta\varphi}$ and $a$ are plotted in the appropriate units to facilitate comparison with the isotropic result for a plasma at the same temperature (left), or at the same entropy density (right). The isotropic result is given in eqs.~\eqn{qisoT} and \eqn{qisos}.
}
\label{qtetayT}
\end{center}
\end{figure}
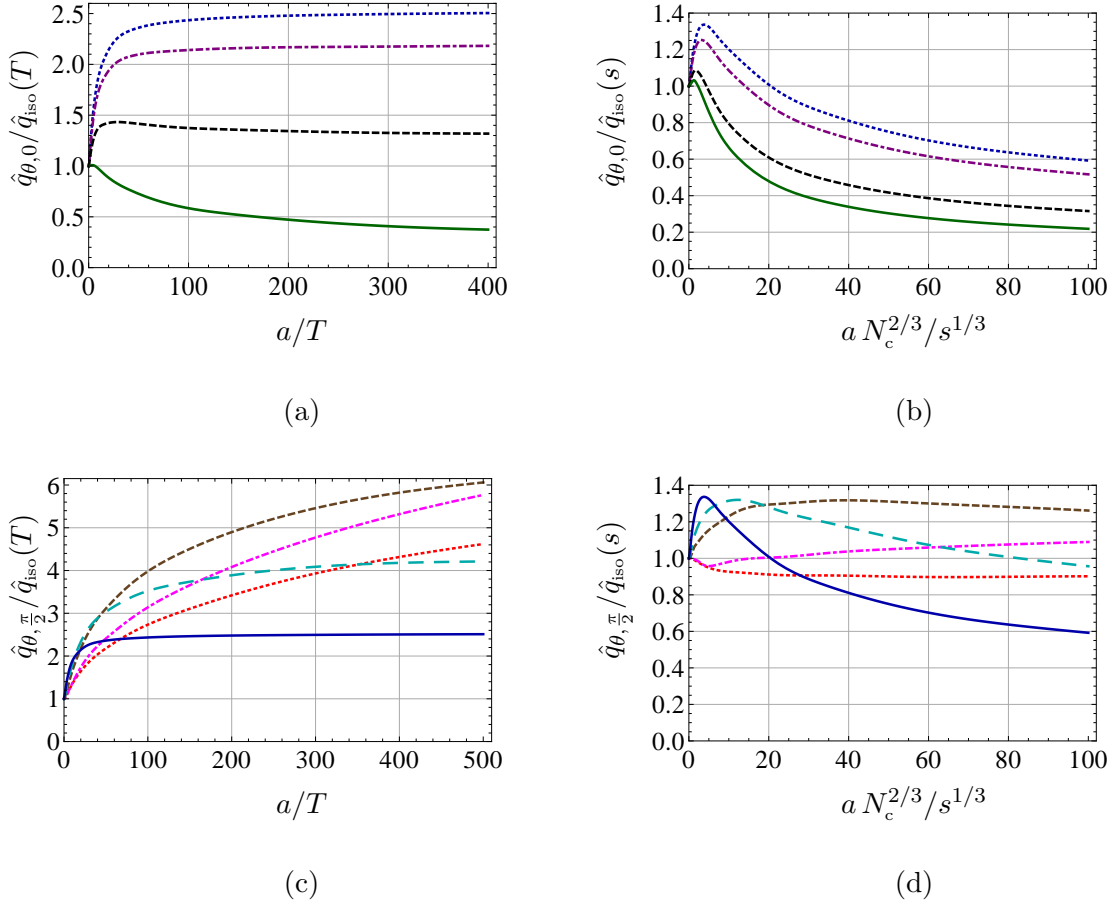

\begin{figure}[htb]
\begin{center} 
\begin{tabular}{cc}
\setlength{\unitlength}{1cm} 
\hskip -1.3cm
\includegraphics[width=9cm,height=5cm]{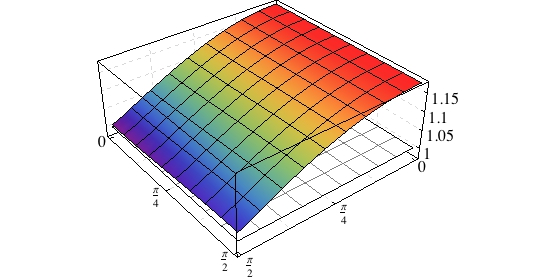}
&  \hskip -1.2cm
\includegraphics[width=9cm,height=5cm]{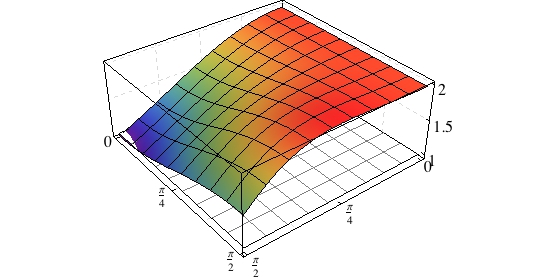}
\\
(a) & (b)\\ &\\
\hskip -1.3cm
\includegraphics[width=9cm,height=5cm]{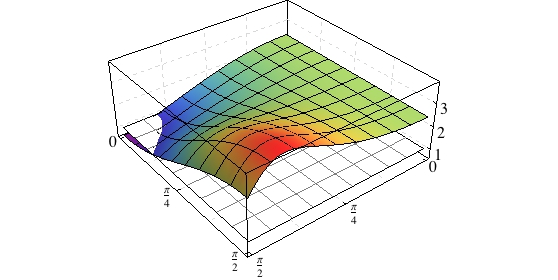}
&  \hskip -1.2cm
\includegraphics[width=9cm,height=5cm]{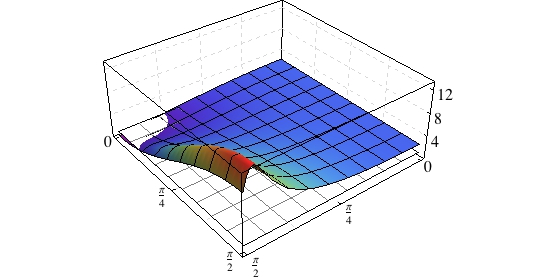}
\\ (c) & (d)
 \begin{picture}(0,0)
  \put(-365,85){\rotatebox{90}
  {$\hat{q}_{\theta, \varphi}/\qh_\mt{iso}(T)$}} 
  \put(-130,85){\rotatebox{90}
  {$\hat{q}_{\theta, \varphi}/\qh_\mt{iso}(T)$}}
   \put(-365,265){\rotatebox{90}
  {$\hat{q}_{\theta, \varphi}/\qh_\mt{iso}(T)$}} 
  \put(-130,265){\rotatebox{90}
  {$\hat{q}_{\theta, \varphi}/\qh_\mt{iso}(T)$}}
     \put(-225,35){$\theta$}
     \put(-328,45){$\varphi$}
     \put(10,35){$\theta$}
     \put(-92,45){$\varphi$}
      \put(-225,210){$\theta$}
     \put(-328,220){$\varphi$}
      \put(10,210){$\theta$}
     \put(-92,220){$\varphi$}
        \end{picture}
     \end{tabular}
\caption{
Jet quenching parameter for a quark moving along an arbitrary direction within the $xz$-plane as a function of the angles $\theta$ and $\varphi$ and for anisotropies $a/T=$ 1.38 (a), 12.2 (b),  86 (c),  3380 (d).
$\qh_{\theta, \varphi}$ is plotted in the appropriate units to facilitate comparison with the isotropic result for a plasma at the same temperature. The isotropic result is given in eq.~\eqn{qisoT}.}
\label{3DplotsT}
\end{center}
\end{figure}

\begin{figure}[h!!]
\begin{center} 
\begin{tabular}{cc}
\setlength{\unitlength}{1cm} 
\hskip -1.3cm
\includegraphics[width=9cm,height=5cm]{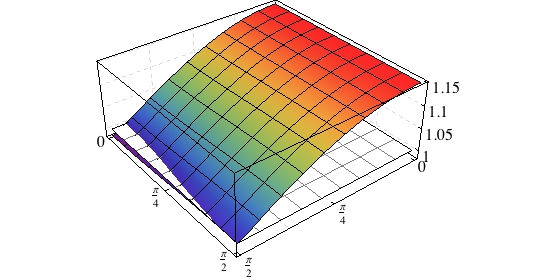}
&  \hskip -1.2cm
\includegraphics[width=9cm,height=5cm]{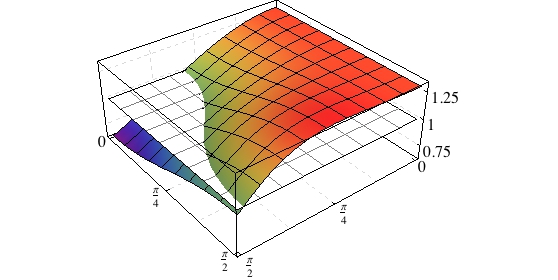}
\\
(a) & (b)\\ &\\
\hskip -1.3cm
\includegraphics[width=9cm,height=5cm]{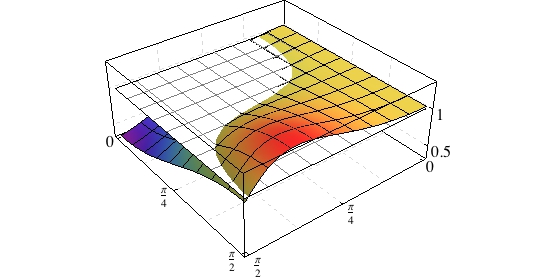}
&  \hskip -1.2cm
\includegraphics[width=9cm,height=5cm]{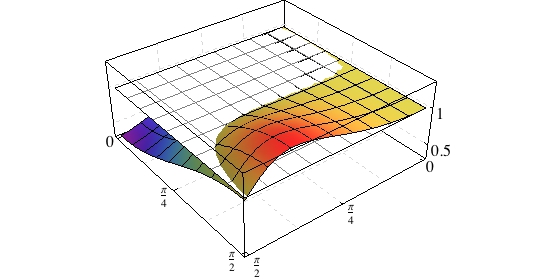}
\\ (c) & (d)\\ &\\
\hskip -1.3cm
\includegraphics[width=9cm,height=5cm]{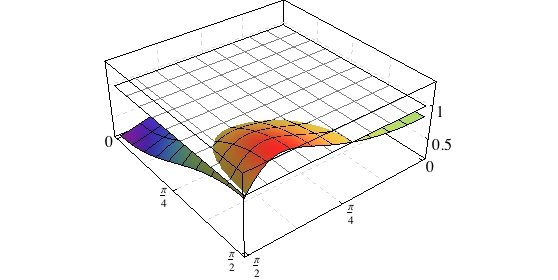}
&  \hskip -1.2cm
\includegraphics[width=9cm,height=5cm]{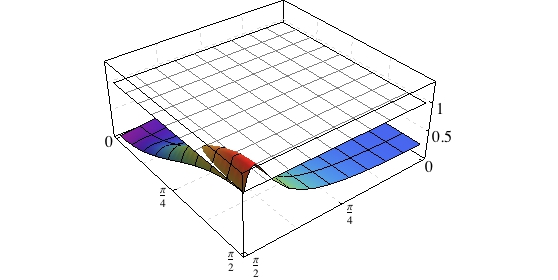}
\\ (e) & (f)
 \begin{picture}(0,0)
  \put(-365,440){\rotatebox{90}
  {$\hat{q}_{\theta, \varphi}/\qh_\mt{iso}(s)$}} 
  \put(-130,440){\rotatebox{90}
  {$\hat{q}_{\theta, \varphi}/\qh_\mt{iso}(s)$}}
  \put(-365,85){\rotatebox{90}
  {$\hat{q}_{\theta, \varphi}/\qh_\mt{iso}(s)$}} 
  \put(-130,85){\rotatebox{90}
  {$\hat{q}_{\theta, \varphi}/\qh_\mt{iso}(s)$}}
   \put(-365,265){\rotatebox{90}
  {$\hat{q}_{\theta, \varphi}/\qh_\mt{iso}(s)$}} 
  \put(-130,265){\rotatebox{90}
  {$\hat{q}_{\theta, \varphi}/\qh_\mt{iso}(s)$}}
     \put(-225,35){$\theta$}
     \put(-328,45){$\varphi$}
     \put(10,35){$\theta$}
     \put(-92,45){$\varphi$}
      \put(-225,210){$\theta$}
     \put(-328,220){$\varphi$}
      \put(10,210){$\theta$}
     \put(-92,220){$\varphi$}
      \put(-225,390){$\theta$}
     \put(-328,397){$\varphi$}
      \put(10,390){$\theta$}
     \put(-92,397){$\varphi$}
        \end{picture}
     \end{tabular}
\caption{
Jet quenching parameter for a quark moving along an arbitrary direction within the $xz$-plane as a function of the angles $\theta$ and $\varphi$ and for anisotropies $a N_\mt{c}^{2/3}/s^{1/3}=0.80$ (a), 6.24 (b),  18.2 (c), 20.2 (d), 35.5 (e),  928 (f).
$\qh_{\theta, \varphi}$ is plotted in the appropriate units to facilitate comparison with the isotropic result for a plasma at the same entropy density. The isotropic result is given in eqn.~\eqn{qisos}.}
\label{3Dplotss}
\end{center}
\end{figure}

Starting from the ansatz above it is straightforward to obtain the Nambu-Goto action (\ref{ng}). However, the resulting expression is quite lengthy and we will not write it down explicitly. As in the previous sections, we can use the fact that the action does not depend explicitly on $Z^+$, $X$, and $Y$. This allows us to express the derivatives with respect to $u$ of these embedding functions in terms of three constants of motion, which we call $\Pi_{+}$, $\Pi_X$, and $\Pi_Y$. We are only interested in the limit in which these quantities are small. In this limit we find
\bea
 (z^+)' & = & c_{++}\Pi_+ + \frac{1}{\sin\varphi} c_{+X}\Pi_X+{\cal O}\left(\Pi^2\right)\,,
 \label{zplusprime}
\\ [5pt]
 X' & = &  \frac{1}{\sin\varphi}c_{X+}\Pi_+ +\frac{1}{\sin^2\varphi} c_{XX}\Pi_X+{\cal O}\left(\Pi^2\right)\,,
 \label{Xprime}
\\[5pt]
 Y' &  = & \frac{1}{\cos^2\varphi}  c_{YY}\Pi_Y+{\cal O}\left(\Pi^2\right)\,,
 \label{yprime}
\eea
with
\bea
c_{++} &\equiv& \frac{1}{\sqrt{2}}\,\frac{u^2(\cf\cb(\cos^2\theta+\ch \sin^2\theta)-\ch)}{\cf\cb\ch\sqrt{\cf(\sin^2\theta+\ch \cos^2\theta-\cf\cb)}} \,, \\[6pt]
c_{XX} &\equiv&\frac{\sqrt{2}\,u^2(\sin^2\theta+\ch \cos^2\theta)}{\ch\sqrt{\cf(\sin^2\theta+\ch \cos^2\theta-\cf\cb)}}\,,
\\[6pt]
c_{+X}&=& c_{X+}\equiv \frac{u^2(\ch-1)\sin\theta\cos\theta}{\ch\sqrt{\cf(\sin^2\theta+\ch \cos^2\theta-\cf\cb)}}\,,\\[6pt]
c_{YY}&\equiv&\frac{\sqrt{2}\,u^2}{\sqrt{\cf(\sin^2\theta+\ch \cos^2\theta-\cf\cb)}}\,.
\eea
An argument analogous to that in Sec.~\ref{1} shows that the denominators in these expressions only vanish at the horizon in the small-$\Pi$ limit.
The endpoints of the string are not separated in the $z^+$-direction, so we must have \mbox{$\int dz^+=0$}. Integrating (\ref{zplusprime}) then gives \be
\Pi_+=-\frac{1}{\sin\varphi}\frac{\int_0^{u_\mt{H}}du \, c_{+X}}{\int_0^{u_\mt{H}}du \, c_{++}}\Pi_X  +{\cal O}\left(\Pi^2\right)\,.
\label{Piplus}
\ee
This result can now be used in the integration of eqn.~(\ref{Xprime}) to obtain $\Pi_X$:
\bea
\Pi_X=\frac{\ell}{2}\frac{\sin^2\varphi\, \int_0^{u_\mt{H}}du \, c_{++}}{\left(\int_0^{u_\mt{H}}du \, c_{++}\right)\left(\int_0^{u_\mt{H}}du \, c_{XX}\right)
-\left(\int_0^{u_\mt{H}}du \, c_{+X}\right)^2}+{\cal O}\left(\Pi^2\right)\,.
\label{PiX}
\eea
Similarly, integrating (\ref{yprime}) yields 
\be
\Pi_Y = \frac{\ell}{2}\frac{\cos^2\varphi}{ \int_0^{u_\mt{H}}du \, c_{YY}}+{\cal O}\left(\Pi^2\right)\,.
\label{Piy}
\ee
Inserting eqs.~(\ref{zplusprime})-(\ref{yprime}) into the action, expanding to quadratic order in the $\Pi$'s and dropping the leading, 
$\Pi$-independent term we find 
\be
S =
2i\frac{\sqrt{\lambda} L^-}{4\pi}\int_0^{u_\mt{H}} du\, \left[
c_{++}\Pi_+^2+\frac{1}{\sin^2\varphi}c_{XX}\Pi_X^2+\frac{2}{\sin\varphi} c_{+X}\Pi_+\Pi_X+\frac{1}{\cos^2\varphi}c_{YY}\Pi_Y^2
\right] . \,\,\,\,\,\,\,\,\,\,\,\,\,
\ee
With the explicit expressions (\ref{Piplus})-(\ref{Piy}) this reduces to 
\be
S =
2i\frac{\sqrt{\lambda} L^- \ell^2}{16\pi}\Big[
P(\theta) \sin^2\varphi+Q(\theta)\cos^2\varphi\Big]\,,
\ee
with
\be
P(\theta)\equiv
\frac{\int_0^{u_\mt{H}}du \, c_{++}}{\left(\int_0^{u_\mt{H}}du \, c_{++}\right)\left(\int_0^{u_\mt{H}}du \, c_{XX}\right)
-\left(\int_0^{u_\mt{H}}du \, c_{+X}\right)^2}\sac 
Q(\theta)\equiv \frac{1}{\int_0^{u_\mt{H}}du \, c_{YY}}\,.
\,\,\,\,\,\,\,\,\,\,\,\,\,
\ee
Using the prescription \eqn{pres} we finally arrive at
\bea
\hat q_{\theta, \varphi}=\frac{\sqrt{2\lambda}}{\pi}\,\Big[
P(\theta)\sin^2\varphi+Q(\theta)\cos^2\varphi\Big] \,.
\label{hatqgeneric}
\eea
We see  that we have indeed derived the expected relation \eqn{relrel} with 
\be
\qh_{\theta,0} = \frac{\sqrt{2\lambda}}{\pi}\, P(\theta) \sac
\qh_{\theta,\pi/2} = \frac{\sqrt{2\lambda}}{\pi}\, Q(\theta) \,.
\ee
Setting $\theta=0$ in \eqn{hatqgeneric} we recover the previous result \eqn{jetqyteta}. In Fig.~\ref{qtetayT} we have plotted the result (\ref{hatqgeneric}) for $\varphi=0$ and $\varphi=\pi/2$ as a function of the ratios $a/T$  and $a N_\mt{c}^{2/3}/s^{1/3}$ for different values of $\theta$. In Figs.~\ref{3DplotsT} and \ref{3Dplotss} we have plotted the result as a function of $\theta$ and $\varphi$ for several values of $a/T$ and $a N_\mt{c}^{2/3}/s^{1/3}$, respectively. 
Note that when $\theta=0$ (motion along the longitudinal direction) the rotational symmetry in the $xy$-plane implies that the jet quenching parameter is independent of $\vp$. For this reason the blue, dotted curves in Figs.~\ref{qtetayT}(a)-(b) agree with the blue, solid curves in Figs.~\ref{qtetayT}(c)-(d). The red, solid curves in Figs.~\ref{qytetaT} also agree with the red, dotted curves in Figs.~\ref{qtetayT}(c)-(d), since they both correspond to $\theta=\vp=\pi/2$. Similarly, the green, dotted curves in Figs.~\ref{qytetaT} agree with the green, solid curves in Figs.~\ref{qtetayT}(a)-(b), since they both correspond to $\theta=\pi/2, \vp=0$.


\section{Discussion}
The momentum broadening of a highly relativistic parton moving through a non-Abelian plasma is described by the jet quenching parameter $\qh$. We have considered an anisotropic ${\cal N}=4$ SYM plasma in which the $x,y$ directions are rotationally symmetric, but the $z$-direction is not. In the context of heavy ion collisions the latter would correspond to the beam direction, and the former to the transverse plane. The jet quenching parameter depends on the relative orientation between these directions on the one hand, and the direction of motion of the parton and the direction in which the momentum broadening is measured, on the other. This dependence can be parametrized by two angles 
$(\theta,\vp)$, as shown in Fig.~\ref{geom}. We have determined the jet quenching parameter $\hat q_{\theta, \varphi}$  for the most  general orientation and for any anisotropy. Our results are valid in the strong-coupling, large-$\nc$ limit, since we have obtained them by means of the gravity dual \cite{prl,jhep} of the anisotropic ${\cal N}=4$  plasma. The anisotropy is induced by a position-dependent theta term in the gauge theory, or equivalently by a position-dependent  axion on the gravity side. One may therefore wonder how sensitive the conclusions may be to the specific source of the anisotropy. In this respect it is useful to note that the gravity calculation involves only the coupling of the string to the background metric. This means that any anisotropy that gives rise to a qualitatively similar metric (and no Neveu-Schwarz $B$-field) will yield qualitatively similar results for the jet quenching parameter irrespectively of the form of the rest of supergravity fields. 

		\begin{figure}[h!!!]
		\begin{center}
		\begin{tabular}{cc}
		\includegraphics[scale=0.5]{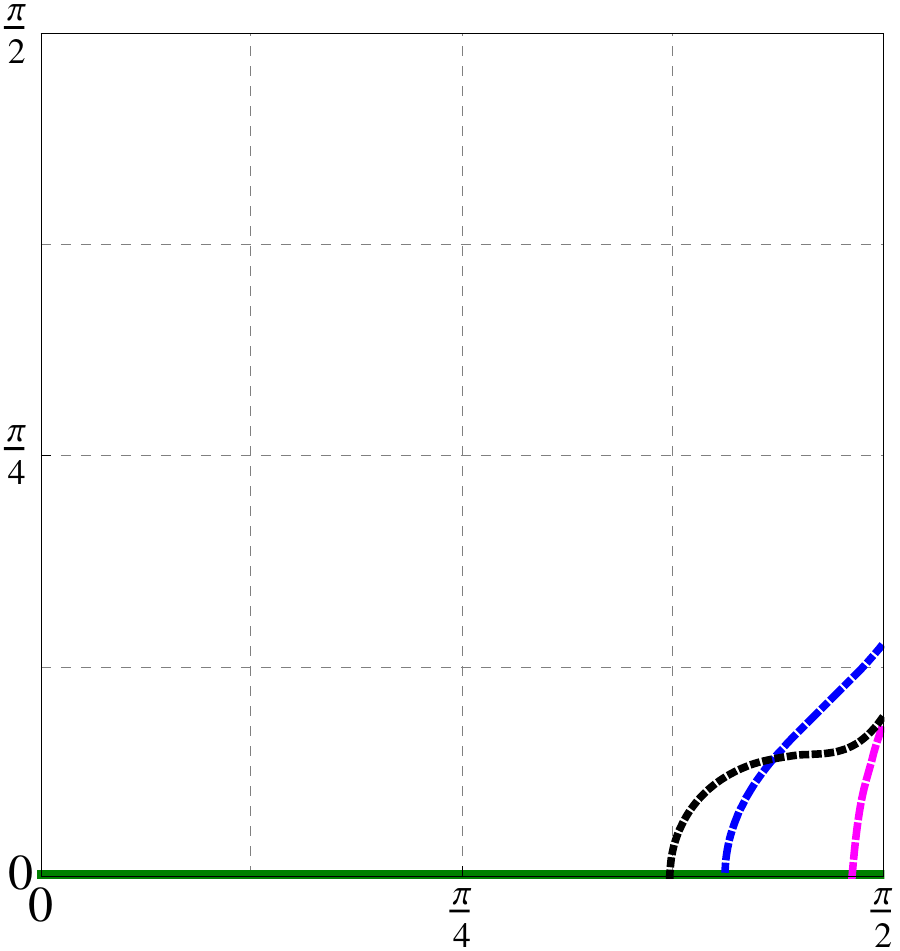}
		\put(-65,-10){$\theta$}
		\put(-145,70){$\varphi$}
		\qquad\qquad &\qquad \qquad 
		\includegraphics[scale=0.5]{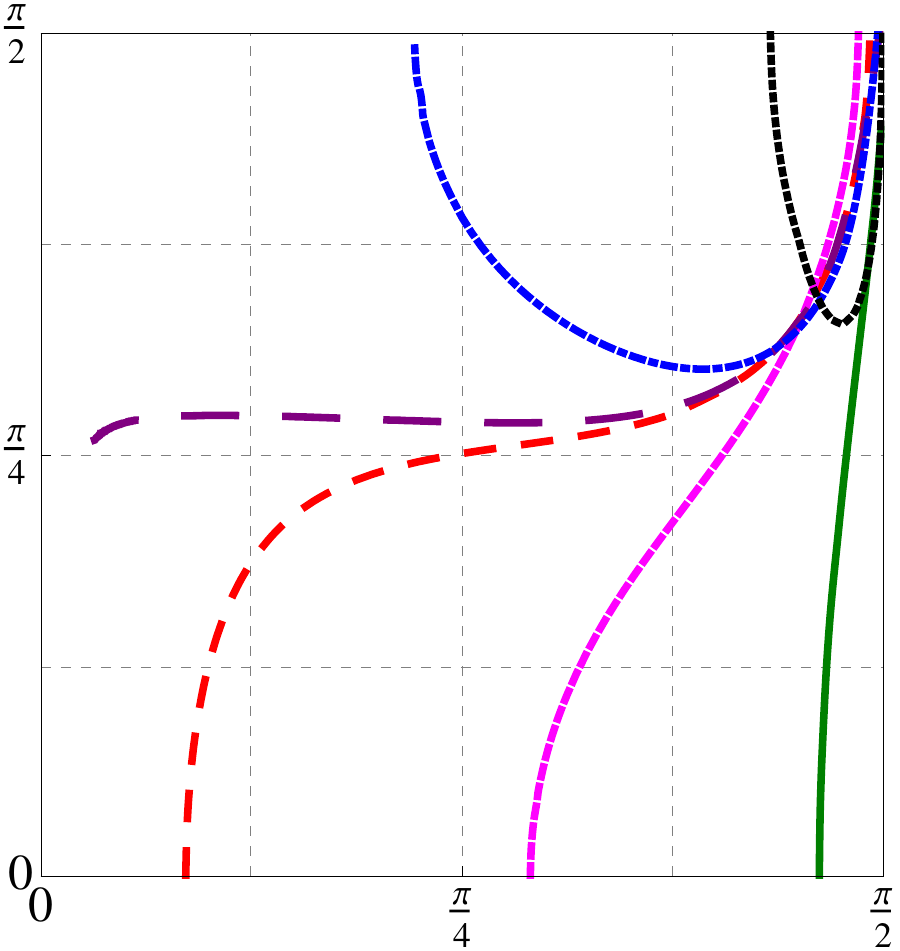}
		\put(-65,-10){$\theta$}
		\put(-145,70){$\varphi$}
		\\ & \\
		(a) \qquad \qquad & \qquad \qquad (b)
		\end{tabular}
		\caption{Above (below) these curves in the $(\theta,\, \varphi)$ plane the jet quenching $\hat q_{\theta, \varphi}$ of the anisotropic plasma is larger (smaller) than the jet quenching of an isotropic plasma at the same temperature (a) or at the same entropy density (b). In (a) the curves correspond to $a/T=1.38$ (green, solid), 12.2 (magenta, dashed), 86 (blue, dot-dashed) and 3380 (black, dotted). In (b) they correspond to $a N_\mt{c}^{2/3}/s^{1/3}=0.80$ (solid, green), 6.24 (magenta, dashed), 18.2 (red, coarsely dashed), 20.2 (purple, very coarsely dashed), 35.5 (blue, dot-dashed) and 928 (black, dotted).
		\label{curves2}
		}
		 \end{center} 
		 \end{figure}		 
For small enough an anisotropy the jet quenching parameter $\hat q_{\theta, \varphi}$ is always larger than that in an isotropic plasma at the same temperature (but different entropy density), regardless of the directions of motion and of momentum broadening. This feature is difficult to appreciate in Figs.~\ref{qzT}(a), \ref{qytetaT}(a), \ref{qtetayT}(a) and  \ref{qtetayT}(c) because of the scale in the horizontal axis, but it can be clearly seen in Fig.~\ref{3DplotsT}(a). Increasing the anisotropy, $\hat q_{\theta, \varphi}$ remains larger than the isotropic value except in a small region close to the $(\theta,\varphi)=(\pi/2,0)$ corner, which we recall corresponds to the momentum broadening along the $y$-direction experienced by a quark propagating along the $x$-axis. 
This region is most clearly shown in Fig.~\ref{curves2}(a), in which we have plotted the curves along which $\hat q_{\theta, \varphi} = \qh_\mt{iso}(T)$, i.e.~the intersections between the two surfaces shown in each of the plots in Fig.~\ref{3DplotsT}. We see that the two regions separated by these curves depend mildly on the value of $a/T$, which varies by more than two orders of magnitude between the magenta, dashed curve ($a/T=12.2$) and the black, dotted curve ($a/T=3380$).

Another interesting feature of the comparison at equal temperature is that, at small $a/T$, $\hat q_{\theta, \varphi}$ is larger for $\theta \simeq 0$, whereas for large $a/T$ the situation gets inverted and $\hat q_{\theta, \varphi}$ becomes larger for $\theta \simeq \pi/2$ (except in the small region close to the $(\theta,\varphi)=(\pi/2,0)$ corner). In other words, at small $a/T$ the momentum broadening is larger for quarks propagating along the beam axis $z$, whereas at large $a/T$ it is larger for quarks propagating in the transverse plane (unless the momentum broadening is measured very close to the orthogonal direction within the transverse plane). Finally, we see that in most of the region where $\hat q_{\theta, \varphi} > \qh_\mt{iso}(T)$, the value of the anisotropic jet quenching parameter increases with $a/T$. This can be seen by noting the scales in the vertical axes in the plots of Fig.~\ref{3DplotsT}, as well as from the slices at constant values of $\theta$ and $\vp$ shown in Figs.~\ref{qzT}(a), \ref{qytetaT}(a), \ref{qtetayT}(a) and \ref{qtetayT}(c).

In contrast, if the comparison is made between plasmas at equal entropy densities (but different temperatures), then the anisotropic jet quenching parameter can be either smaller or larger than its isotropic counterpart for any value of the entropy density, as seen in Fig.~\ref{3Dplotss}. As most clearly shown in Fig.~\ref{curves2}(b), for small $a\nc^{2/3}/s^{1/3}$ the anisotropic jet quenching parameter is greater than the isotropic one except in a small region close to $\theta=\pi/2$, i.e.~for all quarks except those propagating close to the transverse plane. This situation gets progressively inverted as  $a\nc^{2/3}/s^{1/3}$ increases, until for large $a\nc^{2/3}/s^{1/3}$ the anisotropic $\hat q_{\theta, \varphi}$ is only larger than $\qh_\mt{iso}(s)$ near the $(\theta,\varphi)=(\pi/2,\pi/2)$ corner, which we recall corresponds to the momentum broadening along the $z$-direction experienced by a quark propagating along the $x$-direction. Thus we see that when the two plasmas are compared at equal entropy densities, the regions where the anisotropic jet quenching is larger or smaller than the isotropic one depend strongly on the value of the entropy density. 

Also in contrast with the equal-temperature case, at equal entropy densities the value of the jet quenching parameter for almost all orientations of the directions of motion and of momentum broadening decreases as $a\nc^{2/3}/s^{1/3}$ increases. This can be seen from the scale in the vertical axes of Fig.~\ref{3Dplotss}, as well as from the slices at constant values of $\theta$ and $\vp$ shown in Figs.~\ref{qzT}(b), \ref{qytetaT}(b), \ref{qtetayT}(b) and \ref{qtetayT}(d).

One feature that the equal-entropy results share with the equal-temperature ones is that, at small $a\nc^{2/3}/s^{1/3}$,  $\hat q_{\theta, \varphi}$ is larger for $\theta \simeq 0$, whereas for large $a\nc^{2/3}/s^{1/3}$ the situation gets inverted and $\hat q_{\theta, \varphi}$ becomes larger for $\theta \simeq \pi/2$ (except in the small region close to the $(\theta,\varphi)=(\pi/2,0)$ corner). This agreement is of course expected, since the normalizations $\qh_\mt{iso}(T)$ or $\qh_\mt{iso}(s)$ cancel out when comparing the values of $\hat q_{\theta, \varphi}$ for different values of $\theta,\vp$ at constant values of $T$ or $s$. 
 
We will now compare our results to the results for the momentum broadening in the real-world QGP in the presence of anisotropies \cite{roman1,dumitru,dumitru2,baier1}.\footnote{Refs.~\cite{Majumder:2009cf,Mrowczynski4} considered an explicitly time-dependent situation, so we will not attempt a comparison with their results.} This comparison should be interpreted with caution because the sources of anisotropy in the QGP created in a heavy ion collision and in our system are different, and for this reason we will limit our comparison to qualitative features of the results.
In the QGP the anisotropy is dynamical in the sense that it is due to the initial distribution of particles in momentum space, which will evolve in time and eventually become isotropic. In contrast, in our case the anisotropy is due to an external source that keeps the system in an equilibrium anisotropic state that will not evolve in time.  We hope that, nevertheless, our system might provide a good toy model for processes whose characteristic time scale is sufficiently shorter than the time scale controlling the time evolution of the QGP.  
 
The most interesting case to consider in the context of heavy ion collisions is that of a quark propagating within the transverse plane, which we discussed in Sec.~\ref{2}. In this case the momentum broadening along the beam axis, $\ql$, and along the transverse plane, $\qp$, will generically differ. Refs.~\cite{roman1,dumitru,dumitru2} compared these quantities to their isotropic counterpart in a plasma at the same temperature. They found that $\ql \gtrsim \qiso > \qp$, i.e.~that the momentum broadening along the beam axis increases slightly in the presence of anisotropy, whereas the momentum broadening in the transverse plane decreases more significantly. These effects become stronger as the anisotropy grows. These results were suggested as a possible explanation of the asymmetric broadening of jet profiles in the plane of pseudorapidity ($\eta$) and azimuthal angle ($\phi$) \cite{ridge1,ridge2,ridge3,ridge4,ridge5}. 

The calculations in Refs.~\cite{roman1,dumitru,dumitru2} rely on the existence of quasi-particles in the plasma. In contrast, our strongly coupled model possess no quasi-particle excitations. In this model we find that the ordering is indeed  $\ql > \qiso > \qp$ for $a/T \gtrsim 6.35$, but for smaller anisotropies we find that $\ql> \qp > \qiso$.
The latter region is not clearly seen in Fig.~\ref{qytetaT}(a) because of the scale in the horizontal axis, but it is illustrated in Fig.~\ref{3DplotsT}(a), where we see that at $a/T=1.38$ we have   $\hat q_{\theta, \varphi}> \qiso$ for all $\theta,\vp$.  Note that  $a/T \gtrsim 6.35$ is a sizable anisotropy, since the transition between the two limiting behaviours of the entropy density shown in Fig.~\ref{scalings} takes place around $a/T \simeq 3.7$. 

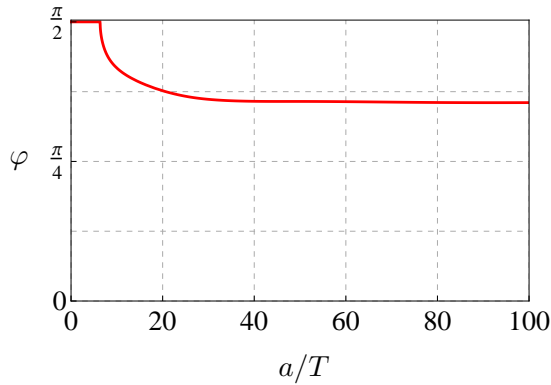
\begin{figure}[t]
\begin{center}
\begin{tabular}{c}
\includegraphics[scale=0.65]{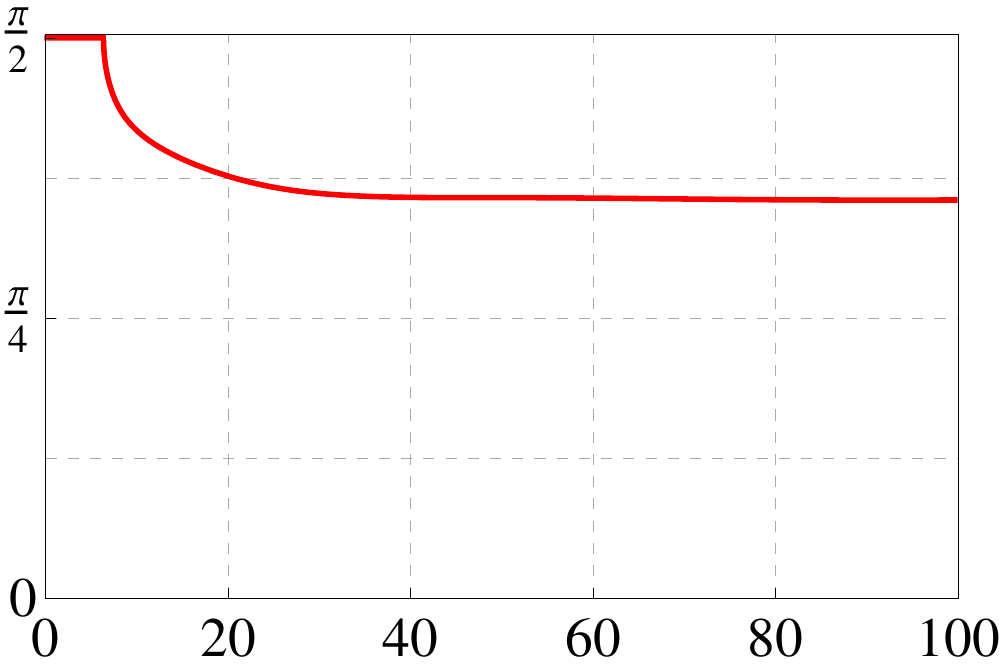}
 \begin{picture}(0,0)
    \put(-206,65){$\vp$}
    \end{picture}
\\
$a/T$    
\end{tabular} 
\caption{Below (above) this curve in the $(a/T,\, \vp)$ plane the jet quenching parameter $\hat q_{x\vp}$ of the anisotropic plasma is larger (smaller) than the jet quenching of the isotropic plasma at the same temperature.
\label{curvepepe}}
\end{center}
\end{figure}

Another difference is that, even for $a/T \gtrsim 6.35$, the most significant effect of the anisotropy is actually on $\ql$, whose increase with $a/T$ is faster than the decrease of $\qp$, as seen in Fig.~\ref{qytetaT}(a). The momentum broadening at an intermediate angle $\vp$ with respect to the transverse plane is given by eqn.~\eqn{jetqyteta}, and this can be smaller or larger than the isotropic value. To illustrate this in Fig.~\ref{curvepepe} we have plotted a curve in the $(a/T,\, \vp)$ plane below (above) which the anisotropic jet quenching parameter is larger (smaller) than its counterpart in an isotropic plasma at the same temperature.  
Finally, we note from Fig.~\ref{qytetaT}(b) that, if the comparison is made at equal entropy densities, then the ordering we find is 
$\qiso > \ql> \qp$ for all values of $a\nc^{2/3}/s^{1/3}$, and moreover the most significant effect in this case is the fast decrease of $\qp$ as $a\nc^{2/3}/s^{1/3}$ increases.

We close by emphasizing one general conclusion of our analysis, namely the fact that whether the jet quenching parameter increases or decreases with respect to its isotropic value depends sensitively on whether the comparison is made at equal temperatures but different entropy densities, or viceversa. This contrasts with our recent calculation of the drag force in the same system \cite{Chernicoff:2012iq}. In that case the comparison between the anisotropic and the isotropic plasmas was relatively insensitive to whether it was done at equal temperatures or at equal entropy densities. This discrepancy is not surprising. The momentum broadening and the drag force are related to each other in the limit $v\to 0$ by the fluctuation-dissipation theorem (see e.g.~\cite{fluct0,fluct1,fluct2} for a discussion in the context of AdS/CFT). However, we have considered the ultra-relativistic limit $v=1$, in which case there is a priori no relation between the momentum broadening and the drag force. 


\begin{acknowledgments}
It is a pleasure to thank Mauricio Martinez, and specially Jorge Casalderrey-Solana, for helpful discussions. 
MC is supported by a postdoctoral fellowship from Mexico's National Council of Science and Technology (CONACyT). 
We acknowledge financial support from 2009-SGR-168, MEC FPA2010-20807-C02-01, MEC FPA2010-20807-C02-02 and CPAN CSD2007-00042 Consolider-Ingenio 2010 (MC, DF and DM), and from DE-FG02-95ER40896 and CNPq (DT).\end{acknowledgments}


\end{document}